\documentclass{elsart5p}

\usepackage{natbib}
\usepackage{graphicx}

\usepackage{amssymb}
\usepackage{url}
\usepackage{balance}

\begin{document}

\begin{frontmatter}



\title{Radio observations of Jupiter-family comets}


\author{Jacques Crovisier\corauthref{cor1}},
\corauth[cor1]{Corresponding author.}
\ead{Jacques.Crovisier@obspm.fr}
\author{Nicolas Biver},
\ead{Nicolas.Biver@obspm.fr}
\author{Dominique Bockel\'ee-Morvan},
\ead{Dominique.Bockelee@obspm.fr}
\author{Pierre Colom}
\ead{Pierre.Colom@obspm.fr}
\address{LESIA, Observatoire de Paris, 5 place Jules Janssen, F-92195
Meudon, France}

\begin{abstract}

Radio observations from decimetric to submillimetric wavelengths are now a basic
tool for the investigation of comets.  Spectroscopic observations allow us i) to
monitor the gas production rate of the comets, by directly observing the water
molecule, or by observing secondary products (e.g., the OH radical) or minor
species (e.g., HCN); ii) to investigate the chemical composition of comets; iii)
to probe the physical conditions of cometary atmospheres: kinetic temperature
and expansion velocity.  Continuum observations probe large-size dust particles
and (for the largest objects) cometary nuclei.

Comets are classified from their orbital characteristics into two separate
classes: i) nearly-isotropic, mainly long-period comets and ii) ecliptic,
short-period comets, the so-called Jupiter-family comets.  These two classes
apparently come from two different reservoirs, respectively the Oort cloud and
the trans-Neptunian scattered disc.  Due to their different history and ---
possibly --- their different origin, they may have different chemical and
physical properties that are worth being investigated.

The present article reviews the contribution of radio observations to our
knowledge of the Jupiter-family comets (JFCs).  The difficulty of such a study
is the commonly low gas and dust productions of these comets.  Long-period,
nearly-isotropic comets from the Oort cloud are better known from Earth-based
observations.  On the other hand, Jupiter-family comets are more easily accessed
by space missions.  However, unique opportunities to observe Jupiter-family
comets are offered when these objects come by chance close to the Earth (like
73P/Schwassmann-Wachmann 3 in 2006), or when they exhibit unexpected outbursts
(as did 17P/Holmes in 2007).

About a dozen JFCs were successfully observed by radio techniques up to now.
Four to ten molecules were detected in five of them.  No obvious evidence for
different properties between JFCs and other families of comets is found, as far
as radio observations are concerned.

\end{abstract}


\end{frontmatter}

\section{Introduction}
\label{sect-intro}

Jupiter-family comets (JFCs, aka \textit{ecliptic comets}) are short-period,
low-inclination comets likely to undergo orbital perturbations by Jupiter.  We
adopt here the definition based on the Tisserand invariant with respect to
Jupiter $T_J$: JFCs are comets for which $2 < T_J < 3$ \citep{levi:1996}.
Special cases are 2P/Encke which slightly exceeds the $T_J = 3$ limit and for
this reason sometimes classified as a \textit{Encke-type} comet, and
29P/Schwassmann-Wachmann 1 with $T_J = 2.99$, alternatively classified as a
\textit{Centaur}.  JFCs are believed to come from the trans-Neptunian scattered
disc.  They contrast with the \textit{nearly isotropic comets}, which comprise
long-period comets as well as short-period comets (the so-called \textit{Halley
type comets}), supposed to come from the Oort cloud.  Comets certainly did not
form in these trans-Neptunian reservoirs.  Their real sites of formation and
their orbital evolution are still highly debated topics.  (For reviews on
cometary families and their dynamical evolution, see \citet{levi:1996,
fern:2008, lowr+:2008, morb:2008, morb+:2008, mars:2008}).  Having different
dynamical histories and --- possibly --- different origins, these distinct
classes of comets may have different chemical and physical properties that are
worth being investigated.

\begin{table*}
\caption{Remote sensing observing conditions for a selection of
comets}
\label{table-remote}
\begin{center}
\begin{tabular*}{\hsize}[]{@{\extracolsep{\fill}}llcccc}
\hline
comet & date & $r_h$ & $\Delta$ & $Q_\mathrm{H_2O}$ & $FM$ \\
      &      & [AU]  & [AU]     & [$10^{28}$ s$^{-1}$] \\
\hline
 \multicolumn{5}{l}{\textit{Nearly-isotropic comets}} \\
C/1986 P1 (Wilson)           & May 1987       & 1.3 & 1.0  &  12 &  12 \\
C/1989 X1 (Austin)           & April 1990     & 1.2 & 0.25 & 2.5 &  10 \\
C/1990 K1 (Levy)             & August 1990    & 1.3 & 0.45 & 25  &  55 \\
C/1996 B2 (Hyakutake)        & March 1996     & 1.1 & 0.11 & 25  & 225 \\
C/1995 O1 (Hale-Bopp)        & April 1997     & 0.9 & 1.4  &1000 & 700 \\
C/1999 S4 (LINEAR)           & July 2000      & 0.77& 0.38 & 10  &  25 \\
C/1999 T1 (McNaught-Hartley) & December 2000  & 1.2 & 1.6  & 10  &   6 \\
C/2001 A2 (LINEAR)           & June 2001      & 1.0 & 0.24 & 10  &  40 \\
C/2000 WM$_1$ (LINEAR)       & December 2001  & 1.2 & 0.32 &  4  &  12 \\
C/2001 Q4 (NEAT)             & May 2004       & 1.0 & 0.32 & 20  &  60 \\
C/2002 T7 (LINEAR)           & May 2004       & 0.8 & 0.27 & 10  &  25 \\
C/2003 K4 (LINEAR)           & December 2004  & 1.4 & 1.2  & 15  &  12 \\
C/2004 Q2 (Machholz)         & January 2005   & 1.2 & 0.35 & 25  &  70 \\
\hline
 \multicolumn{5}{l}{\textit{Halley-type comets}} \\
 1P/Halley                    & January 1986   & 0.7 & 1.5  & 120 &  80 \\
 109P/Swift-Tuttle            & November 1992  & 1.0 & 1.3  & 50  &  40 \\
 153P/2002 C1 (Ikeya-Zhang)   & April 2002   & 1.0 & 0.40 & 25  &  60 \\
 8P/Tuttle                    & January 2008   & 1.1  & 0.25 & 3   & 12 \\
\hline
 \multicolumn{5}{l}{\textit{Jupiter-family comets}} \\
22P/Kopff                    & April 1996     & 1.7  & 0.9  & 3.5 & 4 \\
21P/Giacobini-Zinner         & October 1998   & 1.2  & 1.0  & 3   & 3 \\
19P/Borrelly                 & September 2001 & 1.36 & 1.47 & 3   & 2 \\
2P/Encke                     & November 2003  & 1.0  & 0.25 & 0.5 & 2 \\
9P/Tempel~1                  & July 2005      & 1.5  & 0.77 & 1   & 1.3 \\
73P/Schwassmann-Wachmann~3   & May 2006       & 1.0  & 0.08 & 2   & 25  \\
17P/Holmes                   & October 2007   & 2.4  & 1.6  & $>100$ & $>60$  \\
67P/Churyumov-Gerasimenko    & Mars 2009      & 1.24 & 1.7  & 1   & 0.6 \\
103P/Hartley~2               & October 2010   & 1.1  & 0.12 & 1.2 & 10 \\
45P/Honda-Mrkos-Pajdu\v{s}\'akov\'a & August 2011    & 1.0  & 0.06 & 0.5 &  8 \\
\hline
\end{tabular*}
\end{center}

Date is for best observing conditions. \\  
$r_h$ = distance to Sun and $\Delta$ = distance to Earth at that date. \\
$Q_\mathrm{H_2O}$ = water production rate at that date.  \\
$FM = Q_\mathrm{H_2O}/\Delta$ = \textit{figure of merit}, roughly
proportional to the signal of cometary molecules.
\end{table*}

JFCs and other comets are far from being equally well observed.  To show this
for Earth-based observations, we will use the \textit{figure of merit} parameter
$FM = Q_\mathrm{H_2O}/\Delta$, where $Q_\mathrm{H_2O}$ is the water production
rate in units of $10^{28}$ s$^{-1}$ and $\Delta$ is the distance to the observer
in AU. (Note that this parameter differs from the \textit{figure of merit}
introduced by \citet{mumm+:2002}, which includes a dependency on the distance to
the Sun.)  This parameter is roughly proportional to the expected signal
intensity (for comets at heliocentric distances of the order of 1~AU), and
allows us to evaluate and compare the observability of comets.
Table~\ref{table-remote} gives the figures of merit for Earth-based observations
of recent long-period comets, as well as for recent and future returns of
short-period comets.  One can see that unexpected, long-period comets from the
nearly isotropic class of comets and Halley-type comets offer much better
opportunities than short-period comets.  Indeed, the two best comets in the last
twenty years were C/1996 B2 (Hyakutake) and C/1995 O1 (Hale-Bopp); unprecedented
spectroscopic observations leaded to the identification of many molecules for
the first time in these two objects \citep{bock+:2005}.

All JFCs are comets with low water production rates $Q_\mathrm{H_2O}$ of at most
a few 10$^{28}$~s$^{-1}$ at perihelion.  The best observing conditions occur for
these comets which make a close approach to the Earth (i.e., for $\Delta$
significantly smaller than 1~AU).  This was recently the case for
73P/Schwassmann-Wachmann 3 (minimum geocentric distance $\Delta = 0.08$ AU in
May 2006) and it will also happen for 103P/Hartley~2 in the near future ($\Delta
= 0.12$ AU in October 2010).  But the \textit{figure of merit} of such JFCs is
still lower than that of some long-period comets for which $FM$ could exceed 50.
Other unique opportunities also occur when JFCs exhibit unexpected outbursts
during which the gas production rate may be increased by several orders of
magnitude.  This was recently the case for 17P/Holmes, but such events are rare.

On the other hand, short-period comets, with their predicted returns, are the
only practicable targets for space missions, which are not yet versatile enough
to accommodate unexpected comets.  Flybys at a low velocity and rendezvous are
only possible for ecliptic comets, due to the energy limitations of current
space technology.  1P/Halley is the only explored comet which does not belong to
the Jupiter family: the price to pay was a very high flyby velocity ($\approx
70$ km~s$^{-1}$).

Spectroscopy at radio wavelengths is now a basic tool for the
investigation of comets.  It allows us: 

\begin{itemize}

\item to monitor the gas production rate of the comets, by directly observing
the water molecule, or by observing secondary products (e.g., the OH radical) or
minor species (e.g., HCN);

\item to investigate the chemical composition of comets;

\item to probe the temperature of cometary atmospheres by observing
simultaneously several rotational lines of the same molecule;

\item to investigate the expansion velocity and kinematics of cometary
atmospheres from the observation of line shapes.

\end{itemize}

Continuum observations of comets in the radio spectral range are also well
suited for studying the nucleus and the dust coma via their thermal radiation.
Since millimetric and submillimetric radiation can only be efficiently radiated
from large particles ($\gtrsim 1$ mm) which comprise most of the dust mass, this
technique is a useful probe of the total dust mass production and of the size
distribution of the large grains.

The present article reviews the contribution of radio observations to our
knowledge of JFCs.  The outcome of radar experiments \citep{harm+:2005}, which
probe the nucleus and large dust particles, is beyond the scope of this paper.

\section{The monitoring of gas production rates}
\label{sect-monitor}

\subsection{Observations of the OH 18-cm lines}

\begin{table}
\caption{Short-period numbered comets observed at Nan\c{c}ay (OH 18-cm
lines)}
\label{table-nancay}
\begin{center}
\begin{tabular*}{\hsize}[]{@{\extracolsep{\fill}}lccrl}
\hline
Comet                     & peri. year & $^{a)}$ & $T_J$$^{b)}$ & type$^{c)}$ \\
\hline
  1P/Halley                  & 1986 & D  & $-0.61$ & HTC \\  
  2P/Encke                   & 1977 & -- & 3.03 & ETC \vspace{-2mm} \\
                             & 1980 & M  &            \vspace{-2mm} \\
                             & 1994 & M  &            \vspace{-2mm} \\
                             & 1997 & -- &            \vspace{-2mm} \\
                             & 2003 & D  &            \vspace{-2mm} \\
                             & 2007 & D  &            \\
  4P/Faye                    & 2006 & D  & 2.75 & JFC \\
  6P/d'Arrest                & 1982 & M  & 2.71 & JFC \\
  8P/Tuttle                  & 1994 & M  & 1.60 & HTC \vspace{-2mm} \\
                             & 2008 & D  &            \\
  9P/Tempel 1                & 2005 & D  & 2.97 & JFC \\
 15P/Finlay                  & 1995 & -- & 2.62 & JFC \\
 16P/Brooks 2                & 2001 & -- & 2.88 & JFC \\
 17P/Holmes                  & 2007 & D  & 2.86 & JFC \\
 19P/Borrelly                & 1994 & D  & 2.56 & JFC \vspace{-2mm} \\
                             & 2001 & D  &            \\
 21P/Giacobini-Zinner        & 1985 & D  & 2.47 & JFC \vspace{-2mm} \\
                             & 1998 & D  &            \\
 22P/Kopff                   & 1996 & D  & 2.87 & JFC \\
 23P/Brorsen-Metcalf         & 1989 & D  & 1.11 & HTC \\
 24P/Schaumasse              & 1993 & D  & 2.50 & JFC \vspace{-2mm} \\
                             & 2001 & M  &            \\
 26P/Grigg-Skjellerup        & 1982 & -- & 2.81 & JFC \\
 27P/Crommelin               & 1984 & M  & 1.48 & HTF \\
 45P/Honda-Mrkos-Pajdu\v{s}\'akov\'a & 1995 & M  & 2.58 & JFC \vspace{-2mm} \\
                             & 2001 & -- &            \\
 46P/Wirtanen                & 1997 & -- & 2.82 & JFC \vspace{-2mm} \\
                             & 2002 & -- &            \vspace{-2mm} \\
                             & 2008 & D  &            \\
 64P/Swift-Gehrels           & 1981 & -- & 2.49 & JFC \\
 67P/Churyumov-Gerasimenko   & 1982 & M  & 2.75 & JFC \\
 73P/Schwassmann-Wachmann 3  & 1995 & D  & 2.78 & JFC \vspace{-2mm} \\
                             & 2001 & D  &            \vspace{-2mm} \\
                             & 2006 & D  &            \\
 81P/Wild 2                  & 1997 & M  & 2.88 & JFC \\
 96P/Machholz 1              & 2007 & D  & 1.94 & HTC \\
109P/Swift-Tuttle            & 1992 & D  & $-0.27$ & HTC \\
122P/de Vico                 & 1995 & D  & 0.37 & HTC \\
141P/Machholz 2              & 1994 & M  & 2.71 & JFC \\
153P/Ikeya-Zhang             & 2002 & D  & 0.88 & HTC \\
\hline
\end{tabular*}
\end{center}

$^{a)}$ --: no detection; M: marginal detection; D: clear detection.\\
$^{b)}$ Tisserand parameter.\\
$^{c)}$ JFC: Jupiter-family comet; HTC: Halley-type comet; ETC: Encke-type comet. \\
See details on observations in \citet{crov+:2002} and
\url{http://www.lesia.obspm.fr/planeto/cometes/basecom/index.html}.
\end{table}

The OH radical in comets is a photodestruction product of water, the main
constituent of cometary nucleus ices.  The observation of its 18-cm lines allows
us to trace the production rate and the kinematics of water.  These lines were
first observed in comet C/1973 E1 (Kohoutek) at Nan\c{c}ay \citep{bira+:1974}
and Green Bank \citep{turn:1974}.  They were subsequently systematically
observed.  Their excitation process is now fairly well known and the subject of
extensive modelling: the cometary OH radicals act as a weak maser, whose
inversion is governed by solar UV excitation.  Because of coincidence of the OH
exciting UV lines with Fraunhofer lines, the OH maser inversion strongly depends
upon the comet heliocentric velocity \citep{desp+:1981, schl-ahea:1988}.
Quenching by collisions, which is an efficient process in the inner coma of the
most productive comets, must also be taken into account \citep{desp+:1981,
schl:1988, gera:1990}.

\begin{figure}[h]
\centering
\includegraphics[height=\hsize,angle=270,clip]{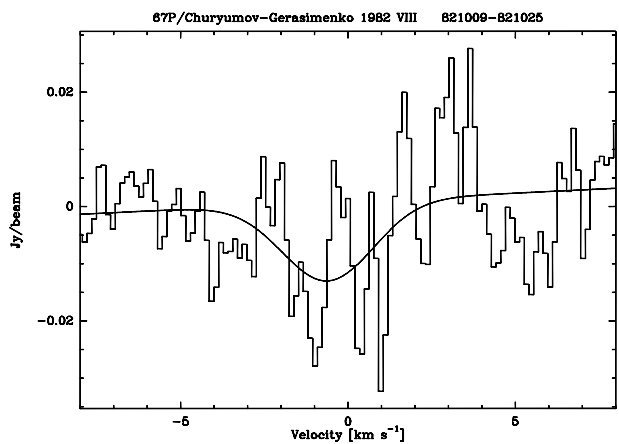}
\caption{Tentative detection of the OH 18-cm lines in comet
67P/Churyumov-Gerasimenko at Nan\c{c}ay in 1982, at $r_h = 1.35$~AU and $\Delta
= 0.50$~AU. The fitted Gaussian corresponds to $Q_\mathrm{OH} = 0.9 \pm 0.2
\times 10^{28}$ s$^{-1}$.  From \citet{crov+:2002}.}
  \label{fig:67p}
\end{figure}

Up to now (beginning of 2008), about 100 comets have been observed with the
Nan\c{c}ay radio telescope \citep[ and in preparation]{desp+:1981, crov+:2002,
crov+:2008-ACM} and with other decimetric radio telescopes.  These observations
include more than 30 passages of Jupiter-family comets (Tables
\ref{table-nancay} and \ref{table-OH}).  In many cases, the weak JFCs were not,
or only marginally, detected (Fig.~\ref{fig:67p}).

\begin{figure}[h]
\centering
\includegraphics[width=\hsize]{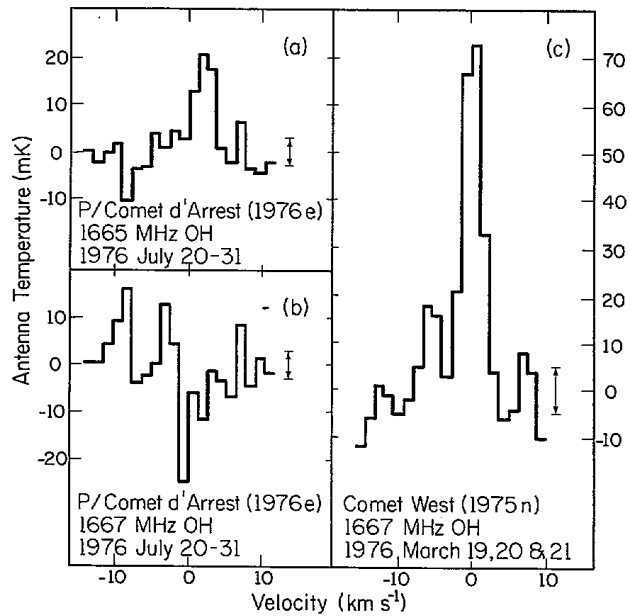}
\caption{The observations at the VRO of the OH 18-cm lines in 6P/d'Arrest (left
panels) and in comet C/1975 V1 West (right panel).  For 6P/d'Arrest, the 1665
MHz line appears in emission and the 1667 MHz is possibly in absorption whereas
excitation models predict absorption for both transitions (heliocentric velocity
--5.6 to --3.1 km~s$^{-1}$).  From \citet{webb-snyd:1977}.}
  \label{fig:VRO_darrest}
\end{figure}

The observation of 6P/d'Arrest at the Vermilion River Observatory in 1976, at
$\Delta \approx 0.2$ AU, was claimed as the first radio detection of a
short-period comet \citep{webb-snyd:1977} (Fig.~\ref{fig:VRO_darrest}).  Was
this detection secure?  The signal-to-noise ratio was poor, and only the 1665
MHz line showed well, in emission whereas the signal is expected in absorption
(Fig.~\ref{fig:VRO_darrest}).  The same comet was only marginally detected at
its following return at Nan\c{c}ay \citep{crov+:2002}.  We consider that the
first reliable detection of a JFC at radio wavelengths was that of
21P/Giacobini-Zinner at its 1985 passage with several radio telescopes
\citep{norr+:1985, galt:1987, gera+:1988, tacc+:1990}.

\begin{table*}
\caption{JFCs comets with observed OH 18-cm lines (Nan\c{c}ay 
excepted)}
\label{table-OH}
\begin{center}
\begin{tabular*}{\hsize}[]{@{\extracolsep{\fill}}lllll}
\hline
Comet                & peri. year & telescope$^{a)}$ & references \\
\hline
 6P/d'Arrest               & 1976 & VRO$^{b)}$     & \citet{webb-snyd:1977} \\
 2P/Encke                  & 1977 & VRO$^{b)}$     & \citet{webb+:1977} \vspace{-2mm} \\  
                           &      & Arecibo$^{b)}$ & Bania, \textit{personal communication} as quoted in \citet{desp+:1981} \\
21P/Giacobini-Zinner       & 1985 & NRAO 140'      & \citet{tacc+:1990} \vspace{-2mm} \\
                           &      & DRAO           & \citet{galt:1987}  \vspace{-2mm} \\
                           &      & VLA$^{b)}$     & \citet{depa+:1991} \vspace{-2mm} \\
                           &      & Parkes         & \citet{norr+:1985} \\
46P/Wirtanen               & 2002 & Arecibo        & \citet{love+:2002} \\
 2P/Encke                  & 2003 & Arecibo        & \citet{howe+:2004} \vspace{-2mm} \\
                           &      & GBT            & idem \\
 9P/Tempel 1               & 2005 & Arecibo        & \citet{howe+:2007a} \vspace{-2mm} \\
                           &      & GBT            & idem \vspace{-2mm} \\
                           &      & Parkes         & \citet{jone+:2006} \\
73P/Schwassmann-Wachmann 3 & 2006 & Arecibo        & \citet{howe+:2007b} \vspace{-2mm} \\
                           &      & GBT            & \citet{love+:2006} \\
\hline
\end{tabular*}
\end{center}

$^{a)}$ 
Arecibo: 300-m telescope at Arecibo, Porto Rico; 
DRAO:    26-m telescope, Dominion Radio Astronomical Observatory, Canada; 
GBT:     100-m telescope at Green Bank, USA;
NRAO     140': 42-m telescope at Green Bank, USA; 
Parkes:  64-m telescope, Australia; 
VLA:     array of 25-m antennas in New Mexico, USA; 
VRO:     36-m telescope at Vermilion River Observatory, Illinois, USA.\\
$^{b)}$ no detection, or doubtful detection.
\end{table*}

\begin{figure}[h]
\centering
\includegraphics[height=\hsize,angle=270,clip]{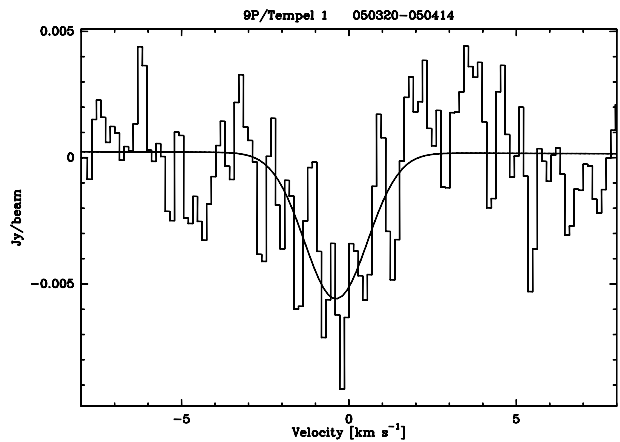}
\caption{Detection of the OH 18-cm lines in comet 9P/Tempel~1 at Nan\c{c}ay in
March--April 2005, at $r_h = 1.76$~AU and $\Delta = 0.79$~AU. The fitted
Gaussian corresponds to $Q_\mathrm{OH} = 0.4 \times 10^{28}$ s$^{-1}$.  From
\citet{bive+:2007-Icarus}.  Compared to Fig.~\ref{fig:67p}, this shows the
improvement in sensitivity of the Nan\c{c}ay radio telescope.}
  \label{fig:9p_nancay}
\end{figure}

\begin{figure}[h]
\centering
\includegraphics[width=\hsize]{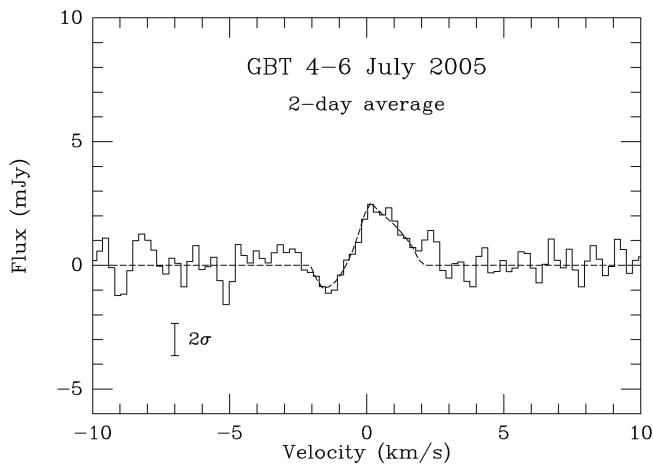}
\caption{The OH 18-cm lines in comet 9P/Tempel~1 observed with the Green Bank
radio telescope just after the impact.  Note the $S$ shape of the line, due to
the Greenstein effect (asymmetric excitation caused by differential velocities
within the coma).  From \citet{howe+:2007a}.}
  \label{fig:9p_GBT}
\end{figure}

The typical 3--$\sigma$ detection limit at Nan\c{c}ay is 6 mJy for a line width
of 2 km~s$^{-1}$ and about 10 hours of integration.  This corresponds to about
$Q_\mathrm{OH} = 0.5 \times 10^{28}$ s$^{-1}$ for a comet at $\Delta \approx
1$~AU at a moment of favourable maser inversion ($\approx 0.40$).  Since the
Nan\c{c}ay telescope can only observe a given target for $\approx 1$~hr per day,
the 10 hours of integration are spread over the same number of days.  The
Arecibo and Green Bank radio telescopes can achieve better limits with longer
integration times per day (Fig.~\ref{fig:9p_GBT}).

\subsection{Observation of the 557~GHz line of water}

Although being the most abundant cometary volatile, water is also one of the
species the most difficult to tackle, because the Earth's atmosphere precludes
its direct observation from the ground (except for weak lines arising from
highly excited rotational or vibrational states).  The $1_{10}$--$1_{01}$
fundamental line of ortho water at 556.936~GHz is expected to be among the
strongest lines of the radio spectrum of comets, but its observation needs to be
made from space.  The first opportunities to observe this line in comets were
provided by the Submillimeter Wave Astronomy Satellite (SWAS) \citep{neuf+:2000}
and the Odin satellite \citep{leca+:2003, bive+:2007-PASS}.  This line is
optically thick: its analysis in order to retrieve column densities and water
production rates requires the coupled modelling of radiative transfer and
molecular excitation \citep{bens-berg:2004, zakh+:2007}.

\begin{figure}[h]
\centering
\includegraphics[width=\hsize]{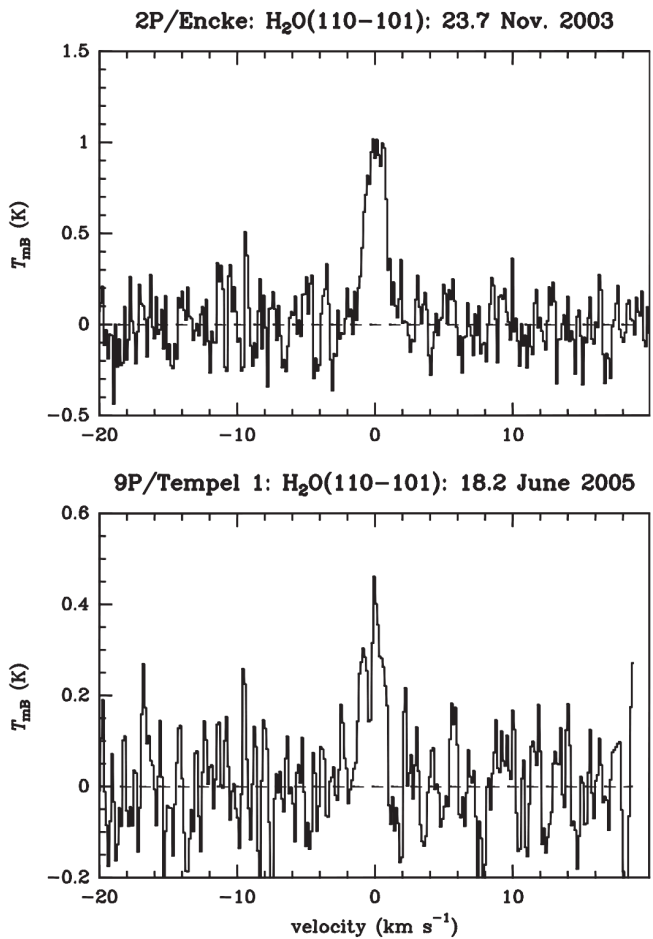}
\caption{Observations of the 557~GHz water line in comets 2P/Encke and
9P/Tempel 1 with Odin.  From \citet{bive+:2007-PASS}.}
  \label{fig:JFC_odin}
\end{figure}

The observations of JFCs with SWAS and Odin are listed in
Table~\ref{table-odin}.  Examples of the water line spectrum are shown in
Fig.~\ref{fig:JFC_odin}.  Unfortunately, these instruments have specific
visibility constraints and cannot observe at small solar elongations.

\begin{table*}
\caption{JFCs observed with SWAS and/or Odin (557~GHz water line)}
\label{table-odin}
\centering
\begin{tabular*}{\hsize}[]{@{\extracolsep{\fill}}llll}
\hline
Comet                & peri. year & satellite & references \\
\hline
19P/Borrelly               & 2001 & Odin & \citet{bock+:2004} \\
29P/Schwassmann-Wachmann 1 &      & Odin & \citet{bive+:2007-PASS} \\
 2P/Encke                  & 2003 & SWAS & \citet{bens-meln:2006-BAAS} \vspace{-2mm} \\
                           &      & Odin & \citet{bive+:2007-PASS} \\
 9P/Tempel 1               & 2005 & SWAS & \citet{bens+:2006-Icarus} \vspace{-2mm} \\
                           &      & Odin & \citet{bive+:2007-Icarus} \\
73P/Schwassmann-Wachmann 3 & 2006 & Odin & \citet{crov+:2006-BAAS, bive+:2008-ACM-73P} \\  
\hline
\end{tabular*}
\end{table*}

Water isotopologues, although observed in some long-period comets (HDO in C/1996
B2 (Hyakutake) and C/1995 O1 (Hale-Bopp) from the ground;
\citet{altw-bock:2003}; H$_2^{18}$O in four comets with Odin), are not yet
detected in JFCs, except for H$_2^{18}$O observed in 73P/Schwassmann-Wachmann 3
by Odin \citep{bive+:2008-ACM-73P}.

In the future, these observations will be pursued and extended with the Herschel
Space Observatory \citep{crov:2005}.  As part of the key programme \textit{Water
and related chemistry in the Solar System} (P.I.\ Paul Hartogh), several
rotational lines of water and its isotopologues are to be observed in JFCs,
including 103P/Hartley~2 at its next passage close to the Earth in October 2010.

The MIRO (\textit{Microwave Instrument for the Rosetta Orbiter}) instrument
\citep{gulk+:2007}, aboard the Rosetta Spacecraft of the European Space Agency,
is also equipped with a radio spectrometer tuned to the $1_{10}$--$1_{01}$ lines
of H$_2$O, H$_2^{18}$O and H$_2^{17}$O (as well as other submillimetric lines of
CO, CH$_3$OH and NH$_3$).  These lines will be observed at a close distance in
67P/Churyumov-Gerasimenko in 2014--2015.

\section{The investigation of chemical composition from millimetric lines}
\label{sect-chemical}

Spectroscopy at radio wavelengths, together with infrared spectroscopy, is
adequate for the investigation of cometary parent molecules \citep[see,
e.g.,][]{bock+:2005}.  The radio technique is very sensitive to trace species
like HCN which have a large dipolar moment.

Table~\ref{table-mm} lists all reported spectroscopic observations of JFCs at
radio wavelengths (excepting water and OH).  This table shows that only few
molecules can be detected in JFCs by radio spectroscopy for standard observing
conditions: four to ten molecules were detected in only five of them.  This is
to be compared to the $\approx 10$ or more molecules detected in several
Oort-cloud comets.  For instance, in 22P/Kopff, only HCN, H$_2$S and CH$_3$OH
could be detected \citep{bive:1997}.  HCN is apparently the easiest molecule to
detect, so that its lines may be used as a proxy for monitoring cometary
activity.  Carbon monoxide has a small dipolar moment which renders the
detection of its rotational lines difficult, despite its presumably large
abundance.  This species could not be detected in the radio in any JFC, except
29P/Schwassmann-Wachmann 1 (Section~\ref{29P}) and 17P/Holmes.  Many more
molecules could be detected in 73P/Schwassmann-Wachmann 3 and 17P/Holmes which
had exceptional observation conditions (Sections \ref{73P} and \ref{17P}).

The diversity of comets from radio observations is discussed in
Section~\ref{sect-family} below.

\begin{table*}
\caption{JFCs observed in spectroscopy with cm, mm or sub-mm ground-based telescopes}
\label{table-mm}
\begin{center}
\begin{tabular*}{\hsize}[]{@{\extracolsep{\fill}}lrlll}
\hline
Comet                        &  perihelion  & telescope$^{a)}$ & molecules  & references \\
\hline  
29P/Schwassmann-Wachmann 1   &              & JCMT & CO                     & \citet{sena-jewi:1994} \vspace{-2mm} \\
                             &              & IRAM & CO                     & \citet{crov+:1995, gunn+:2008} \vspace{-2mm} \\
                             &              & SEST & CO                     & \citet{fest+:2001, gunn+:2002} \\
21P/Giacobini-Zinner         &  5 Sep. 1985 & IRAM & (HCN)$^{b)}$           & \citet{bock+:1987} \vspace{-2mm} \\
                             &              & Effelsberg & (NH$_3$)$^{b)}$  & \citet{bird+:1987} \\
19P/Borrelly                 &  1 Nov. 1994 & IRAM & HCN, CH$_3$OH, H$_2$CO & \citet{bock+:1995, bock+:2004} \vspace{-2mm} \\
                             &              & JCMT & HCN                    & idem \\
45P/Honda-Mrkos-Pajdu\v{s}\'akov\'a  & 26 Dec. 1995 & JCMT & HCN            & \citet{bive:1997, bive+:2002-div} \\
22P/Kopff                    &  2 Jul. 1996 & IRAM & HCN, H$_2$S, CH$_3$OH  & \citet{bive:1997, bive+:2002-div} \\
21P/Giacobini-Zinner         & 21 Nov. 1998 & IRAM & HCN, CS, CH$_3$OH      & \citet{bive+:1999, bive+:2002-div} \vspace{-2mm} \\
                             &              &  CSO & HCN, CH$_3$OH          & idem \vspace{-2mm} \\
                             &              & JCMT & HCN, CS, CH$_3$OH      & idem  \\
52P/Harrington-Abell         & 28 Jan. 1999 & JCMT & (HCN)$^{b)}$, (CO)$^{b)}$ & Biver et al. (unpublished) \\
37P/Forbes                   &   4 May 1999 &  CSO & (HCN)$^{b)}$           & Biver et al. (unpublished) \\
10P/Tempel 2                 &  8 Sep. 1999 &  CSO & HCN, CH$_3$OH          & \citet{bive+:2002-div} \vspace{-2mm} \\
                             &              & JCMT & HCN                    & idem \\
141P/Machholz 2-A            &  9 Dec. 1999 &  CSO & (HCN)$^{b)}$           & Biver et al. (unpublished) \vspace{-2mm} \\
                             &              & JCMT & (HCN)$^{b)}$, (CO)$^{b)}$  & idem \\
 2P/Encke                    &  9 Sep. 2000 &  CSO & (HCN)$^{b)}$           & Biver et al. (unpublished) \\
73P/Schwassmann-Wachmann 3   & 27 Jan. 2001 &  CSO & (HCN)$^{b)}$           & Biver et al. (unpublished) \\
19P/Borrelly                 & 14 Sep. 2001 & IRAM & HCN, CS                & \citet{bock+:2004} \\
67P/Churyumov-Gerasimenko    & 18 Aug. 2002 & IRAM & (CO)$^{b)}$            & \citet{bock+:2004-Capri} \\
 2P/Encke                    & 29 Dec. 2003 & IRAM & HCN, CH$_3$OH          & \citet{crov+:2005} \vspace{-2mm} \\
                             &              &  CSO & HCN, CH$_3$OH          & idem \vspace{-2mm} \\
                             &              & JCMT & HCN                    & \citet{wood+:2003, wood+:2004} \\
 9P/Tempel 1                 &  5 Jul. 2005 & IRAM & HCN, H$_2$S, CH$_3$OH  & \citet{bive+:2007-Icarus} \vspace{-2mm} \\
                             &              &  CSO & CH$_3$OH               & idem \vspace{-2mm} \\
                             &              & JCMT & (HCN)$^{b)}$           & \citet{coul+:2005} \vspace{-2mm} \\ 
                             &              & Mopra& (HCN)$^{b)}$           & \citet{jone+:2006} \vspace{-2mm} \\
                             &              & ATCA & (HCN)$^{b)}$, (CH$_3$OH)$^{b)}$& \citet{jone+:2006} \vspace{-2mm} \\
                             &              & Medicina & (NH$_3$)$^{b)}$    & \citet{tozz+:2007} \\
73P/Schwassmann-Wachmann 3   &7--8 Jun. 2006& IRAM & HCN, CS, H$_2$S, CH$_3$OH,& \citet{bive+:2006-BAAS, bive+:2008-ACM-73P} \vspace{-2mm} \\
                             &              &      & H$_2$CO, CH$_3$CN, HNCO& \vspace{-2mm} \\
                             &              &  CSO & HCN, CH$_3$OH, H$_2$CO & \citet{bive+:2006-BAAS, bive+:2008-ACM-73P, lis+:2008} \vspace{-2mm} \\
                             &              & APEX & HCN, CS, CH$_3$OH      & \citet{bive+:2006-BAAS, bive+:2008-ACM-73P} \vspace{-2mm} \\
                             &              & Kitt Peak & HCN               & \citet{mila+:2006} \vspace{-2mm} \\
                             &              &  SMT & HCN                    & \citet{drah+:2007} \\
17P/Holmes                   &  4 May  2007 & IRAM & HCN, CO, CS, CH$_3$OH, & \citet{bive+:2008-ACM-17P, bock+:2008} \vspace{-2mm} \\
                             &              &      & HNC, H$_2$CO, CH$_3$CN,& \vspace{-2mm} \\
                             &              &      & H$_2$S, SO             & \vspace{-2mm} \\
                             &              & IRAM PdB & HCN, HNC, HC$_3$N, HCO$^+$ & \citet{bois+:2008} \vspace{-2mm} \\
                             &              &  CSO & HCN, CO, CS, CH$_3$OH, & \citet{bive+:2008-ACM-17P} \vspace{-2mm} \\
                             &              &      & HNC, H$_2$CO           & \vspace{-2mm} \\
                             &              & Kitt Peak & HCN, CS, H$_2$S, CH$_3$OH,& \citet{drah+:2007-IAUC, drah+:2008-ACM} \vspace{-2mm} \\
                             &              &           & H$_2$CO           & \vspace{-2mm} \\
                             &              &  SMT & HCN                    & \citet{drah+:2008-IAUC} \\
46P/Wirtanen                 &  2 Feb. 2008 & IRAM & HCN                    & Biver et al. (unpublished) \\
\hline
\end{tabular*}
\end{center}

$^{a)}$ 
APEX:       Atacama Pathfinder Experiment (15-m telescope), Chile; 
ATCA:       Australia Telescope Compact Array;
CSO:        Caltech Submillimeter Observatory 10-m telescope at Mauna Kea, Hawaii, USA; 
Effelsberg: MPIFR 100-m telescope at Effelsberg, Germany;
IRAM:       Institut de radio astronomie millim\'etrique 30-m telescope at Pico Veletta, Spain; 
IRAM PdB:   IRAM interferometer at Plateau de Bure, France;
JCMT:       James Clerk Maxwell Telescope (15 m) at Mauna Kea, Hawaii, USA; 
Kitt Peak:  12-m telescope at Kitt Peak, Arizona, USA; 
Medicina:   32-m telescope, Bologna, Italy;
Mopra:      22-m telescope, Australia;
SEST:       Swedish-ESO Submillimetre Telescope (15 m) at La Silla, Chile; 
SMT:        Sub-Millimeter Telescope (10 m) at Mt Graham, Arizona, USA.\\ 
$^{b)}$     no detection.\\
\end{table*}

\section{Radio continuum measurements}

There is a long history of observations of cometary radio continuum.  The
majority of observations made at centimetric wavelengths were unsuccessful or
matter of debate; see the reviews of \citet{snyd:1982} and
\citet{crov-schl:1991} for early observations.  Even for the productive and
dusty comet C/1995 O1 (Hale-Bopp), only a very weak thermal continuum was
detected at 0.9~cm wavelength, attributed to thermal emission of cm-sized
particles \citep{alte+:1999}.  To our knowledge, the only reported
long-wavelength continuum observation of a JFC is that of comet 6P/d'Arrest at
$\lambda$ = 2.8~cm performed with the NRAO 43-m Green Bank telescope, which was
negative \citep{gibs-hobb:1981}.

\begin{table*}
\caption{JFCs observed in radio continuum (only 17P/Holmes was detected)}
\label{table-cont}
\begin{center}
\begin{tabular*}{\hsize}[]{@{\extracolsep{\fill}}lrlrl}
\hline
Comet                        & perihelion   & telescope$^{a)}$  & frequency & references \\
                             &              &            & [GHz]     &           \\
\hline
6P/d'Arrest                  & 12 Aug. 1976 & NRAO 140'  &  11       & \citet{gibs-hobb:1981} \\
21P/Giacobini-Zinner         &  5 Sep. 1985 & Effelsberg &  24       & \citet{alte+:1986} \\
4P/Faye                      & 16 Nov. 1991 & JCMT       & 375       & \citet{jewi-luu:1992} \\
67P/Churyumov-Gerasimenko    & 18 Aug. 2002 & IRAM       & 250       & \citet{bock+:2004-Capri} \\
17P/Holmes                   &  4 May  2007 & IRAM PdB   &  90       & \citet{bois+:2008} \vspace{-2mm} \\
                             &              & IRAM       & 250       & Altenhoff (personal communication) \vspace{-2mm} \\
                             &              & SMA        &           & Qi (personal communication) \\
\hline
\end{tabular*}
\end{center}

$^{a)}$ See notes to Tables \ref{table-OH} and \ref{table-mm}; SMA:
Submillimeter Array, Hawaii, USA.
\end{table*}

Several long-period comets were detected via their millimetre or submillimetre
thermal radiation, e.g., comets 1P/Halley \citep{alte+:1989}, 109P/Swift-Tuttle
\citep{jewi:1996}, 23P/Brorsen-Metcalf \citep{jewi-luu:1990}, C/1996 B2
(Hyakutake) \citep{jewi-matt:1997, alte+:1999}, C/1995 O1 (Hale-Bopp)
\citep{depa+:1998, alte+:1999, jewi-matt:1999}.  These observations provided
constraints on the total dust mass production rate and, to some extent, on the
particle size distribution from measurements of the spectral index of the dust
emission.  A measure of the size of comet Hale-Bopp's nucleus was obtained from
observations with the IRAM Plateau de Bure interferometer, thanks to the high
angular resolution of the continuum maps which allowed to estimate the relative
contributions of the dust and nucleus emissions \citep{alte+:1999}.  Similar
observations can be made with a single-dish radio telescope on an inactive, or
low-activity comet.  Indeed, this was achieved with the IRAM 30-m telescope on
the Centaur 67P/Chiron \citep{alte-stum:1995} thanks to its large size (84~km
radius).

Only a few millimetre or submillimetre continuum observations of JFCs were
reported (Table~\ref{table-cont}).  Observations of 21P/Giacobini-Zinner
\citep{alte+:1986}, 4P/Faye \citep{jewi-luu:1992} and 67P/Churyumov-Gerasimenko
\citep{bock+:2004} were not conclusive, a result not surprising given their low
activity level.  Note that 67P was observed as it was at 3 AU from the Sun in
order to support Rosetta science operations, following studies of its dust tail
which suggested an unexpected high mass loss rate of large grains
\citep{full+:2004}.

Typically, a comet should have a \textit{figure of merit} (cf.\
Table~\ref{table-remote}) larger than $\approx 7$ to be detected at 250~GHz, on
the basis of a rms noise of 1~mJy, achievable in three hours of winter
observations with the IRAM 30-m telescope.  Other existing major
(sub)millimetric telescopes have similar performances.

Recently, the huge outburst that comet 17P/Holmes underwent on 24 October 2007
allowed the first successful radio continuum observations of a JFC
(Section~\ref{17P}).

\section{Discussion of specific comets}
\label{sect-specific}

\subsection{2P/Encke}
\label{2P}

2P/Encke is the shortest-period comet, famous for its non-gravitational forces
and its seasonal effects.  Observations of OH in this comet are not easy,
because the water production rate is important only close to perihelion (at a
distance $q = 0.338$ AU from the Sun), when the OH lifetime is short.  The
preliminary report of a detection of OH by \citet{webb+:1977} at the 1977
passage, must be considered as dubious.  Only marginal detections were obtained
at Nan\c{c}ay in 1980 and 1994 \citep{bock+:1981, crov+:2002}.  Confirmed
detections of OH were made at the 2003 and 2007 passages \citep[][Crovisier et
al., in preparation]{howe+:2004}.

\begin{figure}[h]
\centering
\includegraphics[height=0.9\hsize,angle=270,clip]{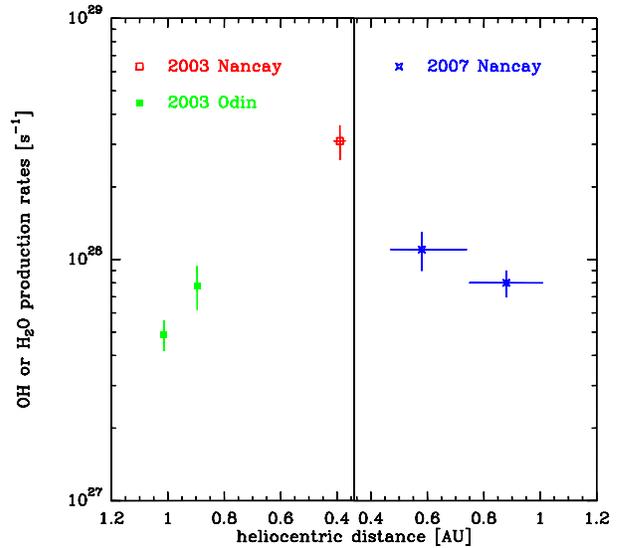}
\caption{The OH and H$_2$O production rates of 2P/Encke, measured at Nan\c{c}ay
(OH) and with Odin (H$_2$O) for different returns, as a function of heliocentric
distance (left is pre-perihelion and right is post-perihelion).  From
\citet[][and in preparation]{crov+:2002} and \citet{bive+:2007-PASS}.}
  \label{fig:2p_prod}
\end{figure}

Fig.~\ref{fig:JFC_odin} shows the water spectrum observed by Odin in 2P/Encke.
Fig.~\ref{fig:2p_prod} shows the OH and water production rates measured for
different passages of 2P/Encke.  HCN and CH$_3$OH could be detected at the 2003
passage.  The methanol abundance is observed to be relatively high: about 4\%
relative to water.

\subsection{21P/Giacobini-Zinner}
\label{21P}

21P/Giacobini-Zinner is the archetype of carbon chain-depleted comets (see
Section~\ref{sect-family}).  It was observed at its 1998 passage as part of a
multi-wavelength campaign.  Only HCN, CS and CH$_3$OH were then detected
\citep{bive+:1999}.  A methanol abundance [CH$_3$OH]/[H$_2$O] = 1.5\% was
observed, which is amongst the lowest abundances observed in comets for this
specie \citep{bive+:2002-div}.  This depletion is confirmed by infrared
observations and is also observed for other volatiles \citep{weav+:1999,
mumm+:2000}.

\subsection{19P/Borrelly}
\label{19P}

This comet was observed at its 1994 passage and in the frame of a
multi-wavelength campaign in support to the flyby of the comet on 22 Sept.
2001 by \textit{Deep Space 1} \citep{sode+:2002}.

$Q_\mathrm{OH}$ was $\approx 2.5 \times 10^{28}$ s$^{-1}$ at $r_h \approx 1.4$
AU at both returns.  Odin observed $Q_\mathrm{H_2O} = 3.8\pm0.5 \times 10^{28}$
s$^{-1}$ at the time of the flyby ($r_h = 1.36$~AU).  Only HCN and CS could be
detected at IRAM in 2001, whereas CH$_3$OH and H$_2$CO were detected at the 1994
passage \citep{bock+:2004}.  The HCN and OH lines were asymmetric and
blueshifted, showing asymmetric outgassing towards the Sun.

\subsection{9P/Tempel 1}
\label{9P}

\begin{figure*}
\centering
\includegraphics[height=0.85\hsize,angle=270,clip]{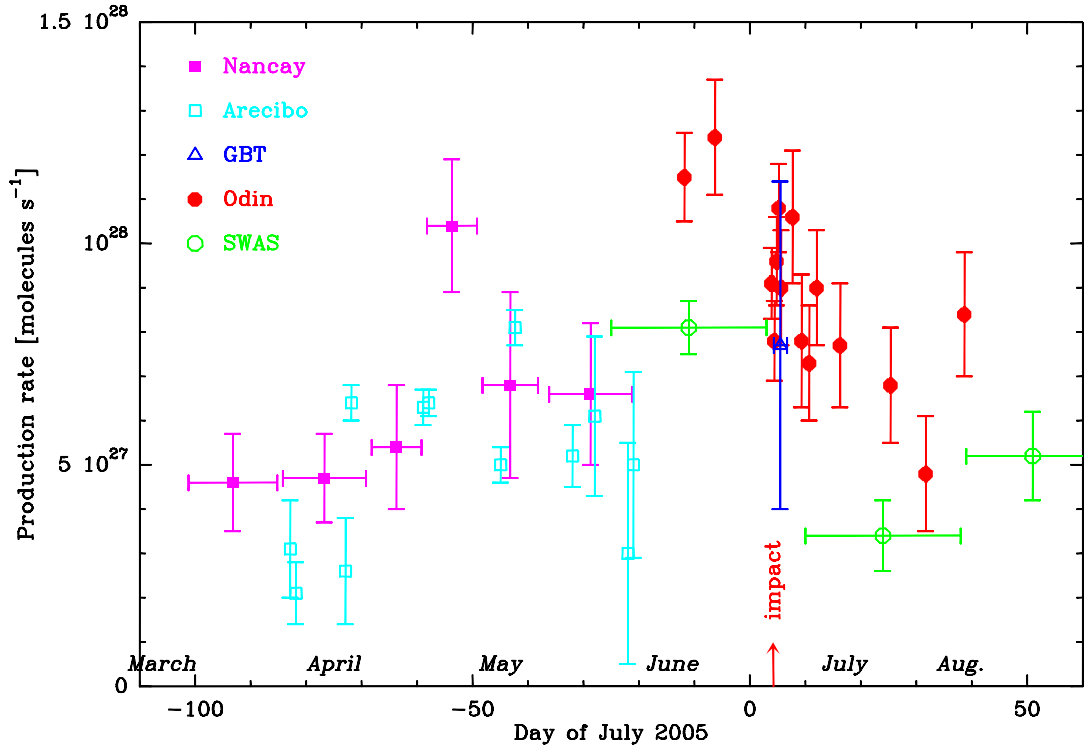}
\caption{The OH and H$_2$O production rates measured in 9P/Tempel 1 from
observations at Nan\c{c}ay, Arecibo, Green Bank and with SWAS and Odin.  The OH
production rates were derived using the OH inversion model of
\citet{desp+:1981}; they have been multiplied by 1.1 to account for the water
photodissociation yield.  From \citet{bive+:2007-Icarus, howe+:2007a,
bens+:2006-Icarus}.}
  \label{fig:9p-prod}
\end{figure*}

The \textit{Deep Impact} mission flew by comet 9P/Tempel~1 on 4 July 2005.  An
impactor was sent to hit the nucleus at 10.2 km~s$^{-1}$ to excavate pristine
material from the sub-surface \citep{ahea+:2005}.  A multi-wavelength campaign
was organised in support to this mission \citep{meec+:2005}.

In March--April 2005, $Q_\mathrm{OH} \approx 4 \times 10^{27}$ s$^{-1}$ was
measured at Nan\c{c}ay when the comet was at $r_h \approx 1.76$~AU
\citep[Fig.~\ref{fig:9p_nancay};][]{bive+:2007-Icarus}.  The OH lines were also
observed in April--June with the Arecibo telescope \citep{howe+:2007a}.  Around
the time of the impact ($r_h = 1.51$~AU), $Q_\mathrm{H_2O}$ was $\approx
10^{28}$ s$^{-1}$ from the observations of SWAS \citep{bens+:2006-Icarus} and
Odin \citep[Fig.~\ref{fig:JFC_odin};][]{bive+:2007-Icarus}.  The OH observations
were difficult at that time because of the small inversion of the OH maser, but
the 18 cm lines could nevertheless be detected with the Green Bank telescope
\citep[Fig.~\ref{fig:9p_GBT};][]{howe+:2007a} and possibly with the Parkes
telescope \citep{jone+:2006}.  The evolution of the OH and H$_2$O production
rates as measured from all radio observations is plotted in
Fig.~\ref{fig:9p-prod}.  The water lines were also searched for --- but not
detected ---around the time of impact by the MIRO radio spectrometer aboard
Rosetta.  Although MIRO could detect the 557~GHz water line of C/2002 T7
(LINEAR) \citep{gulk+:2007}, the 30-cm telescope of this instrument is only
suited for close up observations.

\begin{figure*}
\centering
\includegraphics[height=0.8\hsize,angle=270,clip]{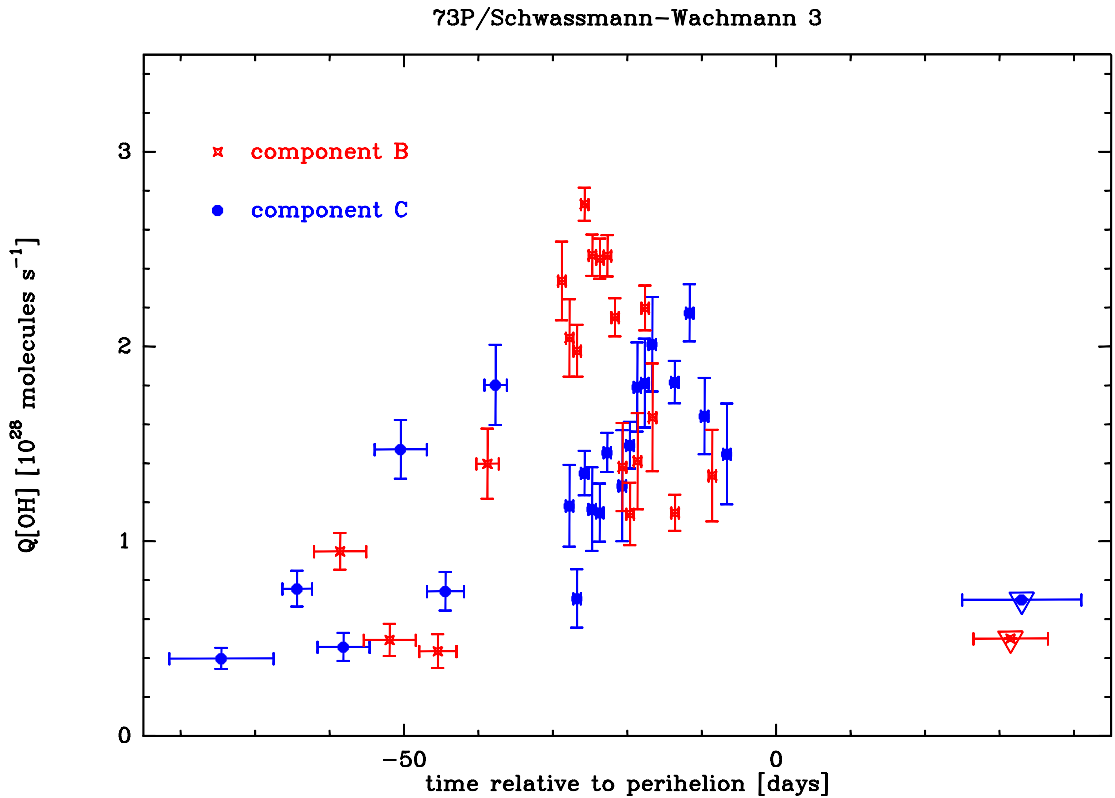}
\caption{Evolutions of the production rates of components B and C of comet
73P/Schwassmann-Wachmann 3, from the OH lines observed at Nan\c{c}ay.  From
\citet{colo+:2006}.}
  \label{fig:SW3}
\end{figure*}

Only HCN, CH$_3$OH and H$_2$S could be detected from other radio spectroscopic
observations \citep[Table~\ref{table-mm};][]{bive+:2007-Icarus}.  These species
show abundances of 0.12, 2.7 and 0.5\% relative to water, respectively, which
are comparable to the mean values observed in many comets.  The variation of the
HCN signal intensity was found to be consistent with the 1.7-day rotation period
of the nucleus.  Post-impact observations did not reveal any significant change
of the outgassing rates or of the relative abundances (except for a possible
increase of the CH$_3$OH outgassing).

\subsection{73P/Schwassmann-Wachmann 3}
\label{73P}

\sloppy The hydroxyl radical was monitored in 73P/Schwassmann-Wachmann 3 at
Nan\c{c}ay at its last three passages \citep{crov+:1996, colo+:2006}.  In 1995,
an unexpected outburst of the OH production was observed, followed by an
increase of the visual brightness.  Then, images showed that the comet had
split.  Since that time, the comet continuously experienced fragmentation
\citep{seka:2005}.  The 2006 passage was especially spectacular, as the comet
made a close approach to Earth at only 0.08 AU in mid-May, just before its
perihelion on 6 June.  The two main fragments, B and C, then both showed water
production rates $\approx 2 \times 10^{28}$ s$^{-1}$ (Fig.~\ref{fig:SW3}).  They
could be the targets of detailed spectroscopic observations with several radio
telescopes \citep[Tables \ref{table-OH}, \ref{table-odin},
\ref{table-mm};][]{bive+:2006-BAAS, bive+:2008-ACM-73P, vill+:2006-BAAS,
drah+:2007, lis+:2008}.

The HCN, CH$_3$CN, HNCO, CH$_3$OH, H$_2$CO, H$_2$S and CS molecules were
detected in addition to OH and H$_2$O. Except for HCN, both fragments show
depletion in volatiles with respect to water, compared to other comets.  The two
fragments B and C were seen to have remarkably similar compositions, suggesting
that the original nucleus of the comet had a homogeneous composition.  Similar
conclusions were drawn from infrared spectroscopic observations
\citep{dell+:2007, disa+:2007, koba+:2007}.

A deep search for hydrogen isocyanide (HNC) was conducted at the CSO in fragment
B \citep{lis+:2008}.  HNC has been observed in a dozen bright comets with ratios
[HNC]/[HCN] ranging from 2 to 15\% and systematically increasing when the
heliocentric distance decreases.  In 73P(B), a low upper limit [HNC]/[HCN]
$\leq$ 1.1\% was found, which is about 7 times lower than the value found in
moderately active comets at similar heliocentric distances.  This puts new
constraints on the still unresolved question of the origin of HNC in comets, and
on the evolution of the [HNC]/[HCN] ratio with heliocentric distance.

\subsection{17P/Holmes}
\label{17P}

The unexpected outburst of 17P/Holmes on 24 October 2007 \citep{gree:2007,
bock:2008-ACM} made it temporarily one of the brightest JFCs
(Table~\ref{table-remote}).  Although water could not be observed at the onset
of the outburst, and although it is difficult to define and compare production
rates for such a variable event, the peak $Q_\mathrm{H_2O}$ possibly exceeded
$10^{30}$ s$^{-1}$ \citep{dell+:2008}.  Many observations could be scheduled on
short notice.  However, at the time of writing, most of the results are not yet
published and it is premature to present a comprehensive review of this
exceptional event.

\begin{figure}[h]
\centering
\includegraphics[width=\hsize]{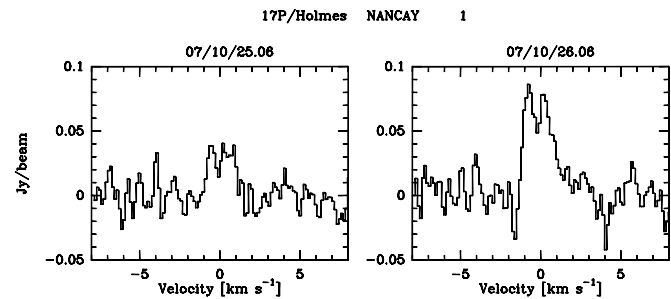}
\caption{The OH spectra of comet 17P/Holmes observed at Nan\c{c}ay just after
its outburst on October 24.2.  From \citet{crov+:2008-ACM}.}
  \label{fig:holmes}
\end{figure}

\begin{figure*}
\centering
\includegraphics[angle=270, width=0.7\hsize]{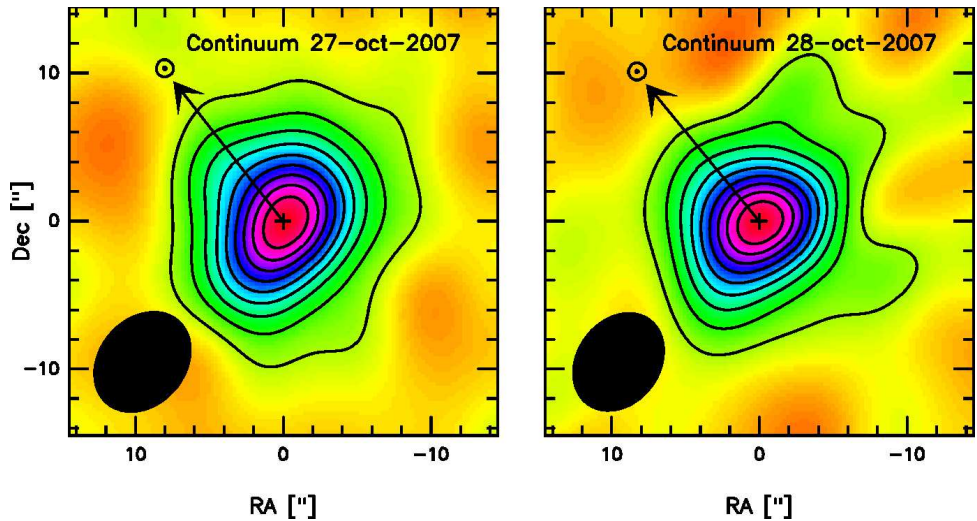}
\caption{The continuum map of 17P/Holmes observed at 90 GHz with the IRAM
interferometer at Plateau de Bure on 27 and 28 October 2007.  From
\citet{bois+:2008}.}
  \label{fig:holmes-cont}
\end{figure*}

Observations could be scheduled at Nan\c{c}ay immediately and the comet was
monitored until 9 November.  The comet was detected initially with a signal
increasing (Fig.~\ref{fig:holmes}); this was followed by a levelling-off and
then a decline in the signal strength.  The line was observed in emission
whereas the OH inversion was expected to be zero or slightly negative.  Thus the
signal was due to thermal emission in a quenched coma.  These observations are
to be compared with the monitoring of the HCN lines made at IRAM, CSO, Kitt Peak
and SMT \citep{bive+:2008-ACM-17P, drah+:2007-IAUC, drah+:2008-IAUC}.

Soon after the outburst of 17P/Holmes, the intensities of the millimetre
molecular lines were comparable to those recorded in comet Hale-Bopp.  It was
thus possible to search for weak signatures of rare isotopes.  The HC$^{15}$N
and H$^{13}$CN $J$(3--2) lines (near 258--259~GHz) were detected at the IRAM
30-m telescope.  Isotopic ratios $^{14}$N/$^{15}$N = $139 \pm 26$ and
$^{12}$C/$^{13}$C = $114 \pm 26$ were derived for HCN, to be compared to the
Earth values of 272 and 89, respectively \citep{bock+:2008}.  This
$^{14}$N/$^{15}$N value in HCN is comparable to that measured in CN for a dozen
comets of different dynamical families, including 17P/Holmes
\citep[e.g.,][]{jehi+:2004, manf+:2005, bock+:2008}.  The same conclusion is
obtained for comet Hale-Bopp after a reanalysis of previous published
measurements of $^{14}$N/$^{15}$N in HCN \citep{jewi+:1997, ziur+:1999,
arpi+:2003, bock+:2008}.  There is some debate on the origin of CN in cometary
atmospheres \citep{fray+:2005}: a similar isotopic composition in HCN and CN is
compatible with HCN being the prime parent of CN. These results also suggest
that JFCs and Oort cloud comets present a similar anomalous nitrogen isotopic
composition in HCN. Important isotopic fractionation of nitrogen should have
taken place at some stage of the Solar System formation under a mechanism which
is still not understood.

Continuum observations were initiated from the millimetre to submillimetre
domains, with the objective to constrain the properties of the dust cloud that
expanded with time.  Continuum maps at $\lambda$ = 3~mm were obtained at the
IRAM Plateau de Bure interferometer two days after the outburst
\citep[Fig.~\ref{fig:holmes-cont};][]{bois+:2008}.  Observations with the IRAM
30-m telescope (Altenhoff, personal communication) and with the Submillimeter
Array (Qi, personal communication) were conducted as well.

\subsection{29P/Schwassmann-Wachmann 1, distant comets and Centaurs}
\label{29P}

At distances $r_h \gtrsim 4$~AU, the nucleus surface equilibrium temperature is
too low to allow a significant sublimation of water ice.  Another mechanism has
to be responsible for cometary activity at large distances from the Sun.
Indeed, the activity of 29P/Schwassmann-Wachmann 1 at $r_h \approx 6$~AU was
found to be governed by the sublimation of carbon monoxide.  This was first
observed from the radio detection of the $J$(2--1) line of CO at the JCMT by
\citet{sena-jewi:1994}.  The observations of radio lines of CO in comets are not
easy: the rotational lines of this molecule, which has a small dipolar moment,
are weak.  On the other hand, infrared and ultraviolet emissions of CO, driven
by fluorescence excited by the Sun, are expected to be even more difficult to
observe in distant comets.

\begin{figure}[h]
\centering
\includegraphics[height=\hsize,angle=270]{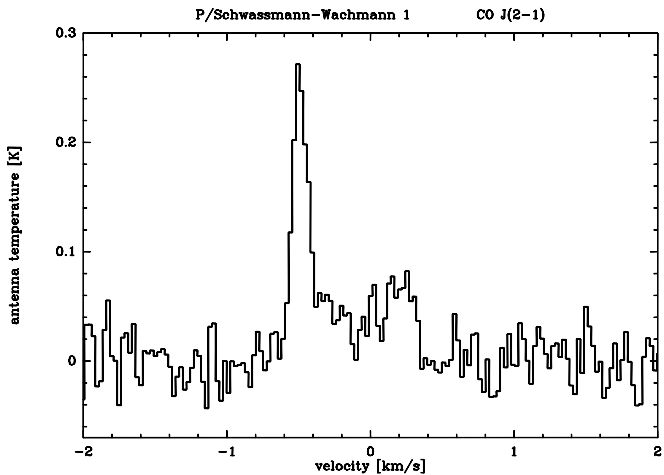}
\caption{The $J$(2--1) line of CO observed in comet 29P/Schwassmann-Wachmann 1
at 6.1 AU from the Sun with the IRAM 30-m telescope.  From \citet{crov+:1995}.}
  \label{fig:SW1}
\end{figure}

29P/Schwassmann-Wachmann 1 is an atypical comet in a nearly circular orbit with
$a = 6.06$~AU, classified either as a JFC or as a Centaur.  Subsequent studies
of CO in this comet were made by \citet{crov+:1995}, \citet{fest+:2001},
\citet{gunn+:2002, gunn+:2008} and \citet{gunn:2003}.  Fig.~\ref{fig:SW1} shows
a representative spectrum of CO in 29P/Schwassmann-Wachmann 1.  The line is
strongly asymmetric with a very narrow component at negative velocities.  The
width of the narrow feature (0.14 km~s$^{-1}$) corresponds to a kinetic
temperature $< 12$ K, which puts stringent constraints on the physical
conditions in the coma.

The inferred CO outgassing is $Q_\mathrm{CO} \approx 4 \times 10^{28}$ s$^{-1}$.
A deep integration on the 557~GHz line of water with Odin in June 2003
yielded a tentative detection (corresponding to $Q_\mathrm{H_2O} \lesssim 2.5
\times 10^{28}$ s$^{-1}$) during the first part of the observation, which could
not be confirmed later \citep{bive+:2007-PASS}.

Searches for an extended source of CO (coming from, e.g., the sublimation of icy
grains) were undertaken by mapping the CO $J$(2--1) line.  Observations at SEST
first suggested such an extended source \citep{fest+:2001}, but were not
confirmed by subsequent observations at IRAM \citep{gunn+:2008}.

\begin{figure*}
\begin{center}
\includegraphics[width=0.7\hsize,angle=270]{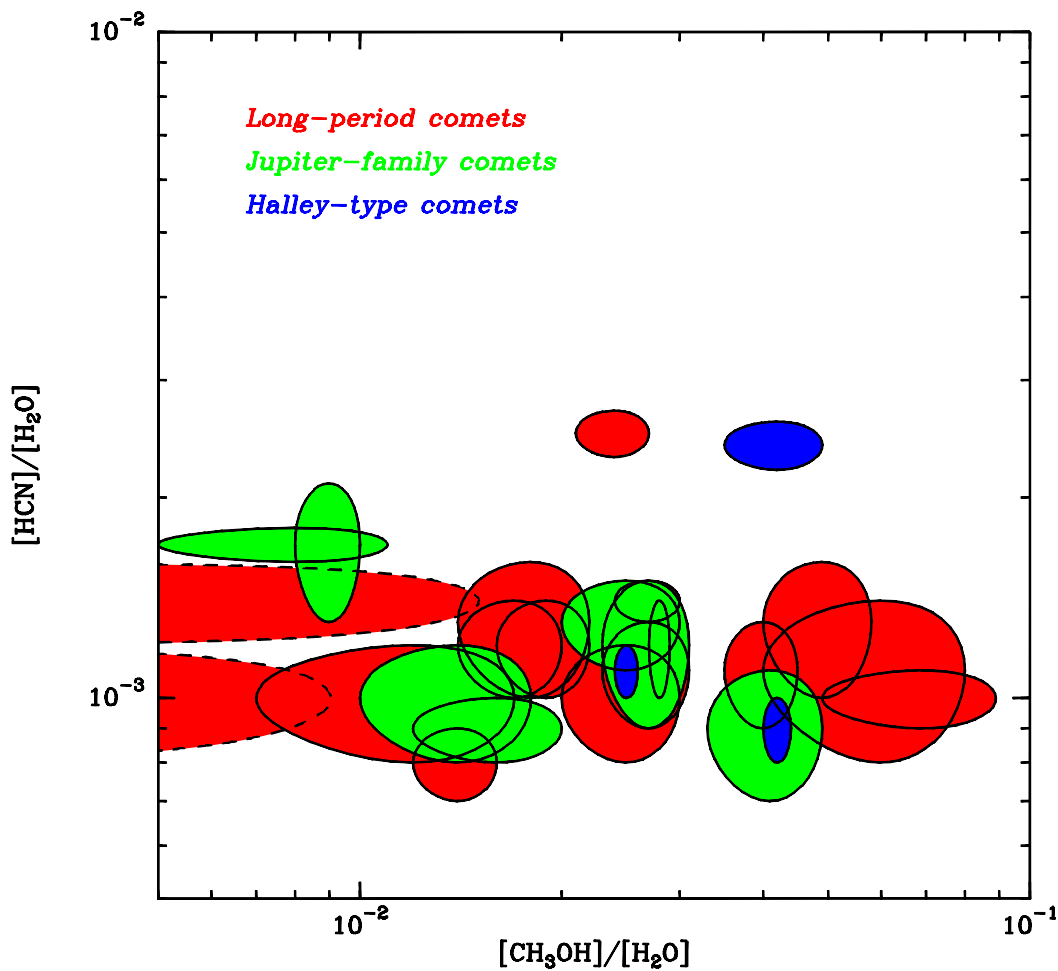}
\caption{The HCN and methanol abundances relative to water for a set of comets
observed by radio spectroscopy.  The sizes of the ellipses represent the errors
of the measurements.  Note that the spread for methanol abundances is much
larger than for HCN. There is no obvious difference between JFCs (in green
in the electronic version) and the other comets.  Adapted from
\citet{crov+:2007-brux}.}
  \label{fig:compo}
\end{center}
\end{figure*}

This discovery of carbon monoxide in a distant comet prompted a search for the
same species in Centaurs.  A detection of CO in 95P/Chiron was claimed by
\citet{woma-ster:1995, woma-ster:1999}.  However, it was not confirmed either in
Chiron or in other Centaurs \citep{raue+:1997, bock+:2001-AA, jewi+:2008}.

The long-period comet C/1995 O1 (Hale-Bopp) was observed by radio spectroscopy
up to $r_h = 14$~AU, allowing us to study the evolution of the sublimation of
water, CO and various other molecules as a function of heliocentric distance
\citep{bive+:2002-moni}.  29P/Schwassmann-Wachmann 1 and C/1995 O1 (Hale-Bopp)
are still the most distant comets in which molecular lines could be detected in
the radio.

\subsection{D1993/F2 Shoemaker-Levy 9 and its collision with Jupiter}
\label{SL9}

To be comprehensive, the case for D1993/F2 Shoemaker-Levy 9 (hereafter SL9)
should be discussed.  SL9 was discovered in late March 1993 after a first
encounter with Jupiter on 7 July 1992 which broke it in multiple fragments and
triggered activity.  The fragments collided with Jupiter on 16--22 July 1994.
Whether SL9 was a bona fide JFC or an asteroid has been the subject of debates
\citep{noll+:1996}.  Indeed, whereas each fragment appeared to be surrounded by
a dust coma, no sign of gas species could be detected before the impacts.
Searches for CO at radio wavelengths at CSO, SEST and IRAM were negative
\citep[M. Festou, D. Jewitt, E. Lellouch, H. Rickman, M. Senay, \textit{personal
communications} as quoted in][] {crov:1996}, as was the case for other searches
for gas in the visible.  The post-impact observations revealed radio lines of
CO, HCN, CS, OCS, possibly H$_2$O \citep{crov:1996}.  Some of these lines
persisted for months or even years, as revealed by the monitoring of the CO, CS
and HCN lines with the JCMT and IRAM 30-m telescopes \citep{more+:2003}.
However, these molecules are not likely to be remnants from the comet.  They
were rather resulting from shock chemistry in the planet's atmosphere.  Their
investigation pertains to the physics and chemistry of Jupiter's atmosphere
rather than to cometary physics \citep{lell:1996}.

\begin{table*}
\caption{Summary of radio observations of JFCs}
\label{table-summary}
\begin{center}
\begin{tabular*}{\hsize}[]{@{\extracolsep{\fill}}lccccccc}
\hline
Comet                               & $FM$   & OH & H$_2$O & HCN & CH$_3$OH & other molecules  & continuum \\
\hline
2P/Encke                            &   2    & X  & X      & X   & X        &         &          \\
4P/Faye                             &  1.3   & X  &        &     &          &         &          \\
6P/d'Arrest                         &   1    & M  &        &     &          &         &          \\
9P/Tempel~1                         &  1.3   & X  & X      & X   & X        & H$_2$S  &          \\
10P/Tempel~2                        &        &    &        & X   & X        &         &          \\
17P/Holmes                          & $>60$  & X  &        & X   & X        & several & X        \\
19P/Borrelly                        &   2    & X  & X      & X   & X        & several &          \\
21P/Giacobini-Zinner                &   3    & X  &        & X   & X        & CS      &          \\
22P/Kopff                           &   4    & X  &        & X   & X        & H$_2$S  &          \\
24P/Schaumasse                      &   2    & X  &        &     &          &         &          \\
29P/Schwassmann-Wachmann 1          & $<0.4$ &    & M      &     &          & CO      &          \\
45P/Honda-Mrkos-Pajdu\v{s}\'akov\'a &   2    & M  &        & X   &          &         &          \\
46P/Wirtanen                        &  0.7   & X  &        & X   &          &         &          \\
67P/Churyumov-Gerasimenko           &  0.6   & M  &        &     &          &         &          \\
73P/Schwassmann-Wachmann 3          &  25    & X  & X      & X   & X        & several &          \\
81P/Wild~2                          &   1    & M  &        &     &          &         &          \\
141P/Machholz 2                     &   2    & M  &        &     &          &         &          \\
\hline
\end{tabular*}
\end{center}

$FM$: figure of merit for best passage (see Table~\ref{table-remote});\\
X: secure detection; M: marginal detection.
\end{table*}

\section{Comparison with other cometary families}
\label{sect-family}

Although the sample is still sparse, there is no obvious correlation between the
dynamical class and the chemical composition of comets \citep{bive+:2002-div,
crov:2007}.  This is shown for the relative abundances of HCN and CH$_3$OH in
Fig.~\ref{fig:compo}.

A taxonomy from narrow-band spectrophotometry in the visible \citep{ahea+:1995}
or from CCD spectroscopy \citep{fink:2006} has been proposed: \textit{typical}
comets and \textit{carbon chain-depleted} comets are distinguished according to
the C$_2$/CN ratio.  Whereas typical comets are present in all dynamical
classes, carbon chain-depleted comets are almost exclusively restricted to JFCs.
Among JFCs well-observed by radio spectroscopy, 2P/Encke, 9P/Tempel~1 and
22P/Kopff are typical, 19P/Borrelly, 21P/Giacobini-Zinner and
73P/Schwassmann-Wachmann 3 are carbon chain-depleted.  These three comets are
also depleted in methanol (with [CH$_3$OH]/[H$_2$O] = 1.7, 1.5 and 0.9\%,
respectively, to be compared to 4, 2.7 and 2.5\%, respectively, for the three
typical comets).  Among the rare non-JFCs which are carbon chain-depleted,
C/1999 S4 (LINEAR) also showed a very low methanol abundance
\citep[{[CH$_3$OH]/[H$_2$O] $\leq$ 1\%};][]{bock+:2001}.

The point of view from infrared observations is discussed by
\citet{disa-mumm:2008}.  The sample is still meager, including 9 comets with
only 2 JFCs.  As for the radio sample, although there are important
comet-to-comet variations of composition, no obvious difference is found among
the various dynamical classes of comets.

\section{Conclusion: Prospects for future observations}

Radio observations of JFCs (see Table~\ref{table-summary} for a summary) are
still sparse.  Up to now, successful radio observations of JFCs have been
limited to a dozen objects, mainly restricted to exceptional comets (comets
which made a close approach to Earth and/or which showed an outburst), or to the
targets of space missions for which a special effort was made.

New radio instruments are to be soon available for the observations of comets
with an increased sensitivity:

\begin{itemize}

\item the Large Millimeter Telescope (LMT), under completion in Mexico
\citep{irvi-schl:2005}; with its 50-m antenna, it will be soon the largest
millimetric telescope of its category;

\item the Herschel Space Observatory, to be launched in 2009 \citep{crov:2005}; 

\item the Atacama Large Millimetre Array (ALMA) in Chili \citep{bive:2005,
bock:2008};

\item the Square Kilometre Array (SKA), which will operate at
decimetric-centimetric wavelengths \citep{butl+:2004};

\item the MIRO instrument on Rosetta, a 30-cm diameter radio telescope to
observe dedicated molecular radio lines in situ in comet
67P/Churyumov-Gerasimenko, as well as the continuum emission of its nucleus
\citep{gulk+:2007}.

\end{itemize}

Specific opportunities to observe JFCs with a high $FM$ will occur in the near
future (Table~\ref{table-remote}):

\begin{itemize}

\item 103P/Hartley~2 in October 2010 (with a flyby of the redirected Deep Impact
spacecraft, renamed as the EPOXI mission);

\item 45P/Honda-Mrkos-Pajdu\v{s}\'akov\'a in August 2011.

\end{itemize}

It will thus be possible to significantly increase the sample of JFCs with known
molecular composition, and to pursue the comparative study of the different
cometary families.

Isotopic measurements in cometary volatiles are also important diagnostics on
the origin of cometary material.  Most measurements have been made using
millimetre or submillimetre spectroscopy.  However, if one excepts the
Jupiter-family comet 17P/Holmes (Section~\ref{17P}), the capabilities of ground
and space-based instrumentation have limited the investigations to a few bright
long-period comets, as reviewed in \citet{altw-bock:2003} and \citet{bock+:2005}
and reported in \citet{bive+:2007-PASS}.  Isotopic ratios could be different in
JFCs and Oort cloud comets if the two populations formed at different places or
different times in the solar nebula.  Though the LMT promises to be very useful
for the study of the molecular composition of JFCs, isotopic investigations with
this telescope will be limited to exceptionally bright comets
\citep{irvi-schl:2002}.  The HIFI instrument of the Herschel Space
Observatory is to provide the opportunity to detect the 1$_{10}$--1$_{01}$ line
of HDO at 509.3~GHz in comet 103P/Hartley 2, and to observe simultaneously
several H$_2$O lines for an accurate determination of the D/H ratio in water.
ALMA will allow the detection of HDO at 465 and 894 GHz in short-period comets
with $FM > 5$, and DCN in comets with $FM > 10$.

\balance

\footnotesize{
\bibliographystyle{elsart-harv}
\bibliography{pss_jfc_radio}

\begin{thebibliography}{129}
\expandafter\ifx\csname natexlab\endcsname\relax\def\natexlab#1{#1}\fi
\expandafter\ifx\csname url\endcsname\relax
  \def\url#1{\texttt{#1}}\fi
\expandafter\ifx\csname urlprefix\endcsname\relax\def\urlprefix{URL }\fi

\bibitem[{{A'Hearn} et~al.(2005){A'Hearn}, {Belton}, {Delamere}, {Kissel},
  {Klaasen}, {McFadden}, {Meech}, {Melosh}, {Schultz}, {Sunshine}, {Thomas},
  {Veverka}, {Yeomans}, {Baca}, {Busko}, {Crockett}, {Collins}, {Desnoyer},
  {Eberhardy}, {Ernst}, {Farnham}, {Feaga}, {Groussin}, {Hampton}, {Ipatov},
  {Li}, {Lindler}, {Lisse}, {Mastrodemos}, {Owen}, {Richardson}, {Wellnitz},
  and {White}}]{ahea+:2005}
{A'Hearn}, M.~F., {Belton}, M.~J.~S., {Delamere}, W.~A., {Kissel}, J.,
  {Klaasen}, K.~P., {McFadden}, L.~A., {Meech}, K.~J., {Melosh}, H.~J.,
  {Schultz}, P.~H., {Sunshine}, J.~M., {Thomas}, P.~C., {Veverka}, J.,
  {Yeomans}, D.~K., {Baca}, M.~W., {Busko}, I., {Crockett}, C.~J., {Collins},
  S.~M., {Desnoyer}, M., {Eberhardy}, C.~A., {Ernst}, C.~M., {Farnham}, T.~L.,
  {Feaga}, L., {Groussin}, O., {Hampton}, D., {Ipatov}, S.~I., {Li}, J.-Y.,
  {Lindler}, D., {Lisse}, C.~M., {Mastrodemos}, N., {Owen}, W.~M.,
  {Richardson}, J.~E., {Wellnitz}, D.~D., {White}, R.~L., 2005. {Deep Impact:
  Excavating comet Tempel 1}. Science 310, 258--264.

\bibitem[{{A'Hearn} et~al.(1995){A'Hearn}, {Millis}, {Schleicher}, {Osip}, and
  {Birch}}]{ahea+:1995}
{A'Hearn}, M.~F., {Millis}, R.~L., {Schleicher}, D.~G., {Osip}, D.~J., {Birch},
  P.~V., 1995. {The ensemble properties of comets: Results from narrowband
  photometry of 85 comets, 1976-1992.} Icarus 118, 223--270.

\bibitem[{{Altenhoff} et~al.(1999){Altenhoff}, {Bieging}, {Butler}, {Butner},
  {Chini}, {Haslam}, {Kreysa}, {Martin}, {Mauersberger}, {McMullin}, {Muders},
  {Peters}, {Schmidt}, {Schraml}, {Sievers}, {Stumpff}, {Thum}, {von Kap-Herr},
  {Wiesemeyer}, {Wink}, and {Zylka}}]{alte+:1999}
{Altenhoff}, W.~J., {Bieging}, J.~H., {Butler}, B., {Butner}, H.~M., {Chini},
  R., {Haslam}, C.~G.~T., {Kreysa}, E., {Martin}, R.~N., {Mauersberger}, R.,
  {McMullin}, J., {Muders}, D., {Peters}, W.~L., {Schmidt}, J., {Schraml},
  J.~B., {Sievers}, A., {Stumpff}, P., {Thum}, C., {von Kap-Herr}, A.,
  {Wiesemeyer}, H., {Wink}, J.~E., {Zylka}, R., 1999. {Coordinated radio
  continuum observations of comets Hyakutake and Hale-Bopp from 22 to 860 GHz}.
  \aap 348, 1020--1034.

\bibitem[{{Altenhoff} et~al.(1989){Altenhoff}, {Huchtmeier}, {Kreysa},
  {Schmidt}, {Schraml}, and {Thum}}]{alte+:1989}
{Altenhoff}, W.~J., {Huchtmeier}, W.~K., {Kreysa}, E., {Schmidt}, J.,
  {Schraml}, J.~B., {Thum}, C., 1989. {Radio continuum observations of Comet
  P/Halley at 250 GHz}. \aap 222, 323--328.

\bibitem[{{Altenhoff} et~al.(1986){Altenhoff}, {Huchtmeier}, {Schmidt},
  {Schraml}, and {Stumpff}}]{alte+:1986}
{Altenhoff}, W.~J., {Huchtmeier}, W.~K., {Schmidt}, J., {Schraml}, J.~B.,
  {Stumpff}, P., 1986. {Radio continuum observations of comet Halley}. \aap
  164, 227--230.

\bibitem[{{Altenhoff} and {Stumpff}(1995)}]{alte-stum:1995}
{Altenhoff}, W.~J., {Stumpff}, P., 1995. {Size estimate of ''asteroid'' 2060
  chiron from 250GHz measurements.} \aap 293, L41--L42.

\bibitem[{{Altwegg} and {Bockel{\'e}e-Morvan}(2003)}]{altw-bock:2003}
{Altwegg}, K., {Bockel{\'e}e-Morvan}, D., 2003. {Isotopic abundances in
  comets}. \ssr 106, 139--154.

\bibitem[{{Arpigny} et~al.(2003){Arpigny}, {Jehin}, {Manfroid},
  {Hutsem{\'e}kers}, {Schulz}, {St{\"u}we}, {Zucconi}, and
  {Ilyin}}]{arpi+:2003}
{Arpigny}, C., {Jehin}, E., {Manfroid}, J., {Hutsem{\'e}kers}, D., {Schulz},
  R., {St{\"u}we}, J.~A., {Zucconi}, J.-M., {Ilyin}, I., 2003. {Anomalous
  nitrogen isotope ratio in comets}. Science 301, 1522--1525.

\bibitem[{{Bensch} and {Bergin}(2004)}]{bens-berg:2004}
{Bensch}, F., {Bergin}, E.~A., 2004. {The pure rotational line emission of
  ortho-water vapor in comets. I. Radiative transfer model}. \apj 615,
  531--544.

\bibitem[{{Bensch} and {Melnick}(2006)}]{bens-meln:2006-BAAS}
{Bensch}, F., {Melnick}, G.~J., 2006. {Submillimeter Wave Astronomy Satellite
  monitoring of the water vaporization rate of seven comets between 1999 and
  2005}. \baas 38, 545.

\bibitem[{{Bensch} et~al.(2006){Bensch}, {Melnick}, {Neufeld}, {Harwit},
  {Snell}, {Patten}, and {Tolls}}]{bens+:2006-Icarus}
{Bensch}, F., {Melnick}, G.~J., {Neufeld}, D.~A., {Harwit}, M., {Snell}, R.~L.,
  {Patten}, B.~M., {Tolls}, V., 2006. {Submillimeter Wave Astronomy Satellite
  observations of comet 9P/Tempel 1 and Deep Impact}. Icarus 184, 602--610.

\bibitem[{{Biraud} et~al.(1974){Biraud}, {Bourgois}, {Crovisier}, {Fillit},
  {G\'erard}, and {Kaz\`es}}]{bira+:1974}
{Biraud}, F., {Bourgois}, G., {Crovisier}, J., {Fillit}, R., {G\'erard}, E.,
  {Kaz\`es}, I., 1974. {OH observation of comet Kohoutek (1973f) at 18 cm
  wavelength}. \aap 34, 163--166.

\bibitem[{{Bird} et~al.(1987){Bird}, {Huchmeier}, {von Kap-herr}, {Schmidt},
  and {Walmsley}}]{bird+:1987}
{Bird}, M.~K., {Huchmeier}, W.~K., {von Kap-herr}, A., {Schmidt}, J.,
  {Walmsley}, C.~M., 1987. {Searches for parent molecules at MPIFR}. In:
  {Irvine}, W.~M., {Schloerb}, F.~P., {Tacconi-Garman}, L.~E. (Eds.), Cometary
  Radio Astronomy, proceedings of an NRAO workshop. NRAO, Green Bank, pp.
  85--87.

\bibitem[{{Biver}(1997)}]{bive:1997}
{Biver}, N., 1997. {Mol\'ecules m\`eres com\'etaires~: Observations et
  mod\'elisations}. PhD~thesis, Universit\'e Paris VII.

\bibitem[{{Biver}(2005)}]{bive:2005}
{Biver}, N., 2005. {Comets with ALMA}. In: {Wilson}, A. (Ed.), The Dusty and
  Molecular Universe. A Prelude to Herschel and ALMA. Vol. 577 of ESA Special
  Publication. pp. 151--156.

\bibitem[{{Biver} et~al.(2006){Biver}, {Bockel{\'e}e-Morvan}, {Boissier},
  {Colom}, {Crovisier}, {Lecacheux}, {Lis}, {Parise}, {Menten}, and {the Odin
  team}}]{bive+:2006-BAAS}
{Biver}, N., {Bockel{\'e}e-Morvan}, D., {Boissier}, J., {Colom}, P.,
  {Crovisier}, J., {Lecacheux}, A., {Lis}, D.~C., {Parise}, B., {Menten}, K.,
  {the Odin team}, 2006. {Comparison of the chemical composition of fragments B
  and C of comet 73P/Schwassmann-Wachmann 3 from radio observations.} \baas 38,
  484--485.

\bibitem[{{Biver} et~al.(2007{\natexlab{a}}){Biver}, {Bockel{\'e}e-Morvan},
  {Boissier}, {Crovisier}, {Colom}, {Lecacheux}, {Moreno}, {Paubert}, {Lis},
  {Sumner}, {Frisk}, {Hjalmarsson}, {Olberg}, {Winnberg}, {Flor\'en},
  {Sandqvist}, and {Kwok}}]{bive+:2007-Icarus}
{Biver}, N., {Bockel{\'e}e-Morvan}, D., {Boissier}, J., {Crovisier}, J.,
  {Colom}, P., {Lecacheux}, A., {Moreno}, R., {Paubert}, G., {Lis}, D.~C.,
  {Sumner}, M., {Frisk}, U., {Hjalmarsson}, {\AA}., {Olberg}, M., {Winnberg},
  A., {Flor\'en}, H.-G., {Sandqvist}, A., {Kwok}, S., 2007{\natexlab{a}}.
  {Radio observations of comet 9P/Tempel 1 before and after Deep Impact}.
  Icarus 187, 253--271.

\bibitem[{{Biver} et~al.(2002){Biver}, {Bockel{\'e}e-Morvan}, {Colom},
  {Crovisier}, {Henry}, {Lellouch}, {Winnberg}, {Johansson}, {Gunnarsson},
  {Rickman}, {Rantakyr{\"o}}, {Davies}, {Dent}, {Paubert}, {Moreno}, {Wink},
  {Despois}, {Benford}, {Gardner}, {Lis}, {Mehringer}, {Phillips}, and
  {Rauer}}]{bive+:2002-moni}
{Biver}, N., {Bockel{\'e}e-Morvan}, D., {Colom}, P., {Crovisier}, J., {Henry},
  F., {Lellouch}, E., {Winnberg}, A., {Johansson}, L.~E.~B., {Gunnarsson}, M.,
  {Rickman}, H., {Rantakyr{\"o}}, F., {Davies}, J.~K., {Dent}, W.~R.~F.,
  {Paubert}, G., {Moreno}, R., {Wink}, J., {Despois}, D., {Benford}, D.~J.,
  {Gardner}, M., {Lis}, D.~C., {Mehringer}, D., {Phillips}, T.~G., {Rauer}, H.,
  2002. {The 1995-2002 long-term monitoring of comet C/1995~O1 (Hale-Bopp) at
  radio wavelength}. Earth Moon and Planets 90, 5--14.

\bibitem[{Biver et~al.(2002)Biver, Bockel\'ee-Morvan, Crovisier, Colom, Henry,
  Moreno, Paubert, Despois, and Lis}]{bive+:2002-div}
Biver, N., Bockel\'ee-Morvan, D., Crovisier, J., Colom, P., Henry, F., Moreno,
  R., Paubert, G., Despois, D., Lis, D.~C., 2002. {Chemical composition
  diversity among 23 comets observed at radio wavelengths}. Earth Moon and
  Planets 90, 323--333.

\bibitem[{{Biver} et~al.(2007{\natexlab{b}}){Biver}, {Bockel{\'e}e-Morvan},
  {Crovisier}, {Lecacheux}, {Frisk}, {Hjalmarson}, {Olberg}, {Flor{\'e}n},
  {Sandqvist}, and {Kwok}}]{bive+:2007-PASS}
{Biver}, N., {Bockel{\'e}e-Morvan}, D., {Crovisier}, J., {Lecacheux}, A.,
  {Frisk}, U., {Hjalmarson}, {\AA}., {Olberg}, M., {Flor{\'e}n}, H.-G.,
  {Sandqvist}, A., {Kwok}, S., 2007{\natexlab{b}}. {Submillimetre observations
  of comets with Odin: 2001--2005}. Planet. Space Scie. 55, 1058--1068.

\bibitem[{{Biver} et~al.(2008{\natexlab{a}}){Biver}, {Bockel{\'e}e-Morvan},
  {Crovisier}, {Lecacheux}, {Lis}, {Boissier}, {Colom}, {Dello-Russo},
  {Flor\'en}, {Frisk}, {Hjalmarson}, {Kwok}, {Menten}, {Moreno}, {Olberg},
  {Parise}, {Paubert}, {Sandqvist}, {Vervack}, {Weaver}, and
  {Winnberg}}]{bive+:2008-ACM-73P}
{Biver}, N., {Bockel{\'e}e-Morvan}, D., {Crovisier}, J., {Lecacheux}, A.,
  {Lis}, D.~C., {Boissier}, J., {Colom}, P., {Dello-Russo}, N., {Flor\'en},
  H.~G., {Frisk}, U., {Hjalmarson}, {\AA}., {Kwok}, S., {Menten}, K., {Moreno},
  R., {Olberg}, M., {Parise}, B., {Paubert}, G., {Sandqvist}, A., {Vervack},
  R.~J., {Weaver}, H.~A., {Winnberg}, A., 2008{\natexlab{a}}. {In-depth
  investigations of the fragmenting comet 73P/Schwassmann-Wachmann 3 at radio
  wavelengths with the Nan\c{c}ay, IRAM, CSO, APEX and Odin radio telescopes}.
  In: Asteroids, Comets, Meteors 2008, Abstract \#8149, Lunar and Planetary
  Institute Contribution No 1405, Houston (CD-ROM).

\bibitem[{{Biver} et~al.(2008{\natexlab{b}}){Biver}, {Bockel{\'e}e-Morvan},
  {Wiesemeyer}, {Crovisier}, {Peng}, {Lis}, {Phillips}, {Boissier}, {Colom},
  {Lellouch}, and {Moreno}}]{bive+:2008-ACM-17P}
{Biver}, N., {Bockel{\'e}e-Morvan}, D., {Wiesemeyer}, H., {Crovisier}, J.,
  {Peng}, R., {Lis}, D., {Phillips}, T., {Boissier}, J., {Colom}, P.,
  {Lellouch}, E., {Moreno}, R., 2008{\natexlab{b}}. {Composition and outburst
  follow-up observations of comet 17P/Holmes at the Nan\c{c}ay, IRAM and CSO
  radio observatories}. In: Asteroids, Comets, Meteors 2008, Abstract \#8186,
  Lunar and Planetary Institute Contribution No 1405, Houston (CD-ROM).

\bibitem[{{Biver} et~al.(1999){Biver}, {Crovisier}, {Davis}, {Despois}, {Lis},
  {Bockel{\'e}e-Morvan}, {Colom}, {G{\'e}rard}, {Matthews}, and
  {Paubert}}]{bive+:1999}
{Biver}, N., {Crovisier}, J., {Davis}, J.~K., {Despois}, D., {Lis}, D.~C.,
  {Bockel{\'e}e-Morvan}, D., {Colom}, P., {G{\'e}rard}, E., {Matthews}, H.~E.,
  {Paubert}, G., 1999. {Coordinated radio observations of comet
  21P/Giacobini-Zinner in October-November 1998}. In: Asteroids, Comets,
  Meteors (book of abstracts). Cornell University, p.~50.

\bibitem[{{Bockel{\'e}e-Morvan}(2008{\natexlab{a}})}]{bock:2008-ACM}
{Bockel{\'e}e-Morvan}, D., 2008{\natexlab{a}}. {Comet 17P/Holmes in outburst}.
  In: Asteroids, Comets, Meteors 2008, Abstract \#8185, Lunar and Planetary
  Institute Contribution No 1405, Houston (CD-ROM).

\bibitem[{{Bockel{\'e}e-Morvan}(2008{\natexlab{b}})}]{bock:2008}
{Bockel{\'e}e-Morvan}, D., 2008{\natexlab{b}}. {Cometary science with ALMA}.
  \apss 313, 183--189.

\bibitem[{{Bockel{\'e}e-Morvan} et~al.(1995){Bockel{\'e}e-Morvan}, {Biver},
  {Colom}, {Crovisier}, {G{\'e}rard}, {Jorda}, {Davies}, {Dent}, {Colas},
  {Despois}, {Paubert}, and {Lamy}}]{bock+:1995}
{Bockel{\'e}e-Morvan}, D., {Biver}, N., {Colom}, P., {Crovisier}, J.,
  {G{\'e}rard}, E., {Jorda}, L., {Davies}, J.~K., {Dent}, B., {Colas}, F.,
  {Despois}, D., {Paubert}, G., {Lamy}, P., 1995. {Visible and radio
  observations of comet 19P/Borrelly}. \baas 27, 1144.

\bibitem[{{Bockel{\'e}e-Morvan} et~al.(2004){Bockel{\'e}e-Morvan}, {Biver},
  {Colom}, {Crovisier}, {Henry}, {Lecacheux}, {Davies}, {Dent}, and
  {Weaver}}]{bock+:2004}
{Bockel{\'e}e-Morvan}, D., {Biver}, N., {Colom}, P., {Crovisier}, J., {Henry},
  F., {Lecacheux}, A., {Davies}, J.~K., {Dent}, W.~R.~F., {Weaver}, H.~A.,
  2004. {The outgassing and composition of comet 19P/Borrelly from radio
  observations}. Icarus 167, 113--128.

\bibitem[{{Bockel\'ee-Morvan} et~al.(2008){Bockel\'ee-Morvan}, {Biver},
  {Jehin}, {Cochran}, {Wiesemeyer}, {Manfroid}, {Hutsem\'ekers}, {Arpigny},
  {Boissier}, {Cochran}, {Colom}, {Crovisier}, {Milutinovic}, {Moreno},
  {Prochaska}, {Ramirez}, {Schulz}, and {Zucconi}}]{bock+:2008}
{Bockel\'ee-Morvan}, D., {Biver}, N., {Jehin}, E., {Cochran}, A.~L.,
  {Wiesemeyer}, H., {Manfroid}, J., {Hutsem\'ekers}, D., {Arpigny}, C.,
  {Boissier}, J., {Cochran}, W., {Colom}, P., {Crovisier}, J., {Milutinovic},
  N., {Moreno}, R., {Prochaska}, J.~X., {Ramirez}, I., {Schulz}, R., {Zucconi},
  J.-M., 2008. {Large excess of heavy nitrogen in both hydrogen cyanide and
  cyanogen from comet 17P/Holmes}. Astrophys. J. 679, L49--L52.

\bibitem[{{Bockel{\'e}e-Morvan}
  et~al.(2001{\natexlab{a}}){Bockel{\'e}e-Morvan}, {Biver}, {Moreno}, {Colom},
  {Crovisier}, {G{\'e}rard}, {Henry}, {Lis}, {Matthews}, {Weaver}, {Womack},
  and {Festou}}]{bock+:2001}
{Bockel{\'e}e-Morvan}, D., {Biver}, N., {Moreno}, R., {Colom}, P., {Crovisier},
  J., {G{\'e}rard}, {\'E}., {Henry}, F., {Lis}, D.~C., {Matthews}, H.,
  {Weaver}, H.~A., {Womack}, M., {Festou}, M.~C., 2001{\natexlab{a}}.
  {Outgassing behavior and composition of comet C/1999 S4 (LINEAR) during its
  disruption}. Science 292, 1339--1343.

\bibitem[{{Bockel\'ee-Morvan} et~al.(1987){Bockel\'ee-Morvan}, {Crovisier},
  {Despois}, {Forveille}, {G\'erard}, {Schraml}, and {Thum}}]{bock+:1987}
{Bockel\'ee-Morvan}, D., {Crovisier}, J., {Despois}, D., {Forveille}, T.,
  {G\'erard}, E., {Schraml}, J., {Thum}, C., 1987. {Molecular obserations of
  comets P/Giacobini-Zinner 1984e and P/Halley 1982i at millimetre
  wavelengths}. \aap 180, 253--262.

\bibitem[{{Bockel\'ee-Morvan} et~al.(1981){Bockel\'ee-Morvan}, {Crovisier},
  {G\'erard}, and {Kaz\`es}}]{bock+:1981}
{Bockel\'ee-Morvan}, D., {Crovisier}, J., {G\'erard}, E., {Kaz\`es}, I., 1981.
  {Observations of the OH radical in comets at 18 cm wavelength}. Icarus 47,
  464--469.

\bibitem[{{Bockel{\'e}e-Morvan} et~al.(2005){Bockel{\'e}e-Morvan}, {Crovisier},
  {Mumma}, and {Weaver}}]{bock+:2005}
{Bockel{\'e}e-Morvan}, D., {Crovisier}, J., {Mumma}, M.~J., {Weaver}, H.~A.,
  2005. {The composition of cometary volatiles}. In: {Festou}, M.~C., {Keller},
  H.~U., {Weaver}, H.~A. (Eds.), Comets II. Univ. Arizona Press, Tucson, pp.
  391--423.

\bibitem[{{Bockel{\'e}e-Morvan}
  et~al.(2001{\natexlab{b}}){Bockel{\'e}e-Morvan}, {Lellouch}, {Biver},
  {Paubert}, {Bauer}, {Colom}, and {Lis}}]{bock+:2001-AA}
{Bockel{\'e}e-Morvan}, D., {Lellouch}, E., {Biver}, N., {Paubert}, G., {Bauer},
  J., {Colom}, P., {Lis}, D.~C., 2001{\natexlab{b}}. {Search for CO gas in
  Pluto, Centaurs and Kuiper Belt objects at radio wavelengths}. \aap 377,
  343--353.

\bibitem[{{Bockel\'ee-Morvan} et~al.(2004){Bockel\'ee-Morvan}, {Moreno},
  {Biver}, {Crovisier}, {Crifo}, {Fulle}, and {Grewing}}]{bock+:2004-Capri}
{Bockel\'ee-Morvan}, D., {Moreno}, R., {Biver}, N., {Crovisier}, J., {Crifo},
  J.-F., {Fulle}, M., {Grewing}, M., 2004. {CO and dust productions in
  67P/Churyumov-Gerasimenko at 3 AU post-perihelion}. In: {Colangeli}, L.,
  {Epifani}, E.~M., {Palumbo}, P. (Eds.), The New Rosetta Targets,
  Observations, Simulations and Instrument Performances. Kluwer Academic
  Publishers, Dordrecht, pp. 25--36.

\bibitem[{{Boissier} et~al.(2008){Boissier}, {Bockel{\'e}e-Morvan}, {Biver,},
  {Crovisier}, {Lellouch}, and {Moreno}}]{bois+:2008}
{Boissier}, J., {Bockel{\'e}e-Morvan}, D., {Biver,}, N., {Crovisier}, J.,
  {Lellouch}, E., {Moreno}, R., 2008. {Interferometric mapping of dust
  continuum, HCN and HNC 3-mm emissions in comet 17P/Holmes at IRAM Plateau de
  Bure}. In: Asteroids, Comets, Meteors 2008, Abstract \#8081, Lunar and
  Planetary Institute Contribution No 1405, Houston (CD-ROM).

\bibitem[{{Butler} et~al.(2004){Butler}, {Campbell}, {de Pater}, and
  {Gary}}]{butl+:2004}
{Butler}, B.~J., {Campbell}, D.~B., {de Pater}, I., {Gary}, D.~E., 2004. {Solar
  System science with SKA}. New Astronomy Review 48, 1511--1535.

\bibitem[{{Colom} et~al.(2006){Colom}, {Crovisier}, {Biver},
  {Bockel{\'e}e-Morvan}, {Boissier}, and {Lecacheux}}]{colo+:2006}
{Colom}, P., {Crovisier}, J., {Biver}, N., {Bockel{\'e}e-Morvan}, D.,
  {Boissier}, J., {Lecacheux}, A., 2006. {Observations of comet
  73P/Schwassmann-Wachmann 3 with the Nan{\c c}ay radio telescope}. In:
  {Barret}, D., {Casoli}, F., {Lagache}, G., {Lecavelier}, A., {Pagani}, L.
  (Eds.), SF2A-2006: Semaine de l'Astrophysique Fran{\c c}aise. pp. 389--393.

\bibitem[{{Coulson} et~al.(2005){Coulson}, {Butner}, {Moriarty-Schieven},
  {Woodney}, {Charnley}, {Rodgers}, {Stuwe}, {Schultz}, {Meech}, {Fernandez},
  and {Vora}}]{coul+:2005}
{Coulson}, I.~M., {Butner}, H.~M., {Moriarty-Schieven}, G., {Woodney}, L.~M.,
  {Charnley}, S.~B., {Rodgers}, S.~D., {Stuwe}, J., {Schultz}, R., {Meech}, K.,
  {Fernandez}, Y., {Vora}, P., 2005. {Deep Impact: Submillimetre spectroscopic
  HCN observations of 9P/Tempel-1 from JCMT}. In: Protostars and Planets V
  conference proceedings. LPI Contribution No. 1286, p. 8524.

\bibitem[{{Crovisier}(1996)}]{crov:1996}
{Crovisier}, J., 1996. {Observational constraints on the composition and nature
  of comet D/Shoemaker-Levy 9}. In: {Noll}, K.~S., {Weaver}, H.~A., {Feldman},
  P.~D. (Eds.), The Collision of Comet Shoemaker-Levy 9 and Jupiter. Cambridge
  University Press, pp. 31--54.

\bibitem[{{Crovisier}(2005)}]{crov:2005}
{Crovisier}, J., 2005. {Comets and asteroids with the Herschel Space
  Observatory}. In: {Wilson}, A. (Ed.), The Dusty and Molecular Universe. A
  Prelude to Herschel and ALMA. Vol. 577 of ESA Special Publication. pp.
  145--150.

\bibitem[{{Crovisier}(2007)}]{crov:2007}
{Crovisier}, J., 2007. {Cometary diversity and cometary families}. In:
  {Celnikier}, L. (Ed.), XVIIIemes Rencontres de Blois: Planetary Science:
  Challenges and Discoveries (in press), arXiv:astro-ph/0703785.

\bibitem[{{Crovisier} et~al.(2006){Crovisier}, {Biver}, {Bockel\'ee-Morvan},
  {Boissier}, {Colom}, {Lecacheux}, {Lis}, {Parise}, {Menten}, and {the Odin
  team}}]{crov+:2006-BAAS}
{Crovisier}, J., {Biver}, N., {Bockel\'ee-Morvan}, D., {Boissier}, J., {Colom},
  P., {Lecacheux}, A., {Lis}, D.~C., {Parise}, B., {Menten}, K., {the Odin
  team}, 2006. {The evolution of the outgassing of fragments B and C of comet
  73P/Schwassmann-Wachmann 3 from radio observations}. \baas 38, 485.

\bibitem[{{Crovisier} et~al.(2007){Crovisier}, {Biver}, {Bockel{\'e}e-Morvan},
  {Boissier}, {Colom}, {Lecacheux}, {Moreno}, {Paubert}, {Lis}, {Sumner},
  {Frisk}, {Hjalmarsson}, {Olberg}, {Winnberg}, {Flor\'en}, {Sandqvist}, and
  {Kwok}}]{crov+:2007-brux}
{Crovisier}, J., {Biver}, N., {Bockel{\'e}e-Morvan}, D., {Boissier}, J.,
  {Colom}, P., {Lecacheux}, A., {Moreno}, R., {Paubert}, G., {Lis}, D.~C.,
  {Sumner}, M., {Frisk}, U., {Hjalmarsson}, {\AA}., {Olberg}, M., {Winnberg},
  A., {Flor\'en}, H.-G., {Sandqvist}, A., {Kwok}, S., 2007. {The chemical
  composition of comet 9P/Tempel 1 from radio observations}. In: {Sterken}, C.,
  {K\"{a}ufl}, H.~U. (Eds.), Deep Impact as a world observatory event:
  synergies in space, time and wavelength. Springer-Verlag, ESO Astrophysics
  Symposia, in press.

\bibitem[{{Crovisier} et~al.(2005){Crovisier}, {Biver}, {Bockel{\'e}e-Morvan},
  {Boissier}, {Colom}, {Moreno}, {Lis}, {Paubert}, {Despois}, {Gunnarsson}, and
  {Weaver}}]{crov+:2005}
{Crovisier}, J., {Biver}, N., {Bockel{\'e}e-Morvan}, D., {Boissier}, J.,
  {Colom}, P., {Moreno}, R., {Lis}, D.~C., {Paubert}, G., {Despois}, D.,
  {Gunnarsson}, M., {Weaver}, H.~A., 2005. {Chemical diversity of comets
  observed at radio wavelengths in 2003-2005}. \baas 37, 646.

\bibitem[{{Crovisier} et~al.(1995){Crovisier}, {Biver}, {Bockelee-Morvan},
  {Colom}, {Jorda}, {Lellouch}, {Paubert}, and {Despois}}]{crov+:1995}
{Crovisier}, J., {Biver}, N., {Bockelee-Morvan}, D., {Colom}, P., {Jorda}, L.,
  {Lellouch}, E., {Paubert}, G., {Despois}, D., 1995. {Carbon monoxide
  outgassing from comet P/Schwassmann-Wachmann 1}. Icarus 115, 213--216.

\bibitem[{{Crovisier} et~al.(1996){Crovisier}, {Bockel\'ee-Morvan}, {Gerard},
  {Rauer}, {Biver}, {Colom}, and {Jorda}}]{crov+:1996}
{Crovisier}, J., {Bockel\'ee-Morvan}, D., {Gerard}, E., {Rauer}, H., {Biver},
  N., {Colom}, P., {Jorda}, L., 1996. {What happened to comet
  73P/Schwassmann-Wachmann 3?} \aap 310, L17--L20.

\bibitem[{{Crovisier} et~al.(2008){Crovisier}, {Colom}, {Biver,}, and
  {Bockel{\'e}e-Morvan}}]{crov+:2008-ACM}
{Crovisier}, J., {Colom}, P., {Biver,}, N., {Bockel{\'e}e-Morvan}, D., 2008.
  {Recent observations of the OH 18-cm lines in comets with the Nan\c{c}ay
  radio telescope}. In: Asteroids, Comets, Meteors 2008, Abstract \#8119, Lunar
  and Planetary Institute Contribution No 1405, Houston (CD-ROM).

\bibitem[{{Crovisier} et~al.(2002){Crovisier}, {Colom}, {G{\'e}rard},
  {Bockel{\'e}e-Morvan}, and {Bourgois}}]{crov+:2002}
{Crovisier}, J., {Colom}, P., {G{\'e}rard}, E., {Bockel{\'e}e-Morvan}, D.,
  {Bourgois}, G., 2002. {Observations at Nan{\c c}ay of the OH 18-cm lines in
  comets. The data base. Observations made from 1982 to 1999}. \aap 393,
  1053--1064.

\bibitem[{{Crovisier} and {Schloerb}(1991)}]{crov-schl:1991}
{Crovisier}, J., {Schloerb}, F.~P., 1991. {The study of comets at radio
  wavelengths}. In: {Newburn}, Jr., R.~L., {Neugebauer}, M., {Rahe}, J. (Eds.),
  IAU Colloq. 116: Comets in the post-Halley era. Vol. 167 of Astrophysics and
  Space Science Library. Kluwer Academic Publishers, Dordrecht, pp. 149--173.

\bibitem[{{de Pater} et~al.(1998){de Pater}, {Forster}, {Wright}, {Butler},
  {Palmer}, {Veal}, {A'Hearn}, and {Snyder}}]{depa+:1998}
{de Pater}, I., {Forster}, J.~R., {Wright}, M., {Butler}, B.~J., {Palmer}, P.,
  {Veal}, J.~M., {A'Hearn}, M.~F., {Snyder}, L.~E., 1998. {BIMA and VLA
  observations of comet Hale-Bopp at 22-115 GHz}. \aj 116, 987--996.

\bibitem[{{de Pater} et~al.(1991){de Pater}, {Palmer}, and
  {Snyder}}]{depa+:1991}
{de Pater}, I., {Palmer}, P., {Snyder}, L.~E., 1991. {A review of radio
  interferometric imaging of comets}. In: {Newburn}, Jr., R.~L., {Neugebauer},
  M., {Rahe}, J. (Eds.), IAU Colloq. 116: Comets in the post-Halley era. Vol.
  167 of Astrophysics and Space Science Library. Kluwer Academic Publishers,
  Dordrecht, pp. 175--207.

\bibitem[{{Dello Russo} et~al.(2007){Dello Russo}, {Vervack}, {Weaver},
  {Biver}, {Bockel{\'e}e-Morvan}, {Crovisier}, and {Lisse}}]{dell+:2007}
{Dello Russo}, N., {Vervack}, R.~J., {Weaver}, H.~A., {Biver}, N.,
  {Bockel{\'e}e-Morvan}, D., {Crovisier}, J., {Lisse}, C.~M., 2007.
  {Compositional homogeneity in the fragmented comet 73P/Schwassmann-Wachmann
  3}. \nat 448, 172--175.

\bibitem[{{Dello Russo} et~al.(2008){Dello Russo}, {Vervack}, {Weaver},
  {Montgomery}, {Deshpande}, {Fern\'andez}, and {Martin}}]{dell+:2008}
{Dello Russo}, N., {Vervack}, R.~J., {Weaver}, H.~A., {Montgomery}, M.~M.,
  {Deshpande}, R., {Fern\'andez}, Y.~R., {Martin}, E.~L., 2008. {The volatile
  composition of comet 17P/Holmes after its extraordinaty outburst}. Astrophys.
  J. 680, 793--802.

\bibitem[{{Despois} et~al.(1981){Despois}, {G\'erard}, {Crovisier}, and
  {Kaz\`es}}]{desp+:1981}
{Despois}, D., {G\'erard}, E., {Crovisier}, J., {Kaz\`es}, I., 1981. {The OH
  radical in comets - Observation and analysis of the hyperfine microwave
  transitions at 1667 MHz and 1665 MHz}. \aap 99, 320--340.

\bibitem[{{DiSanti} et~al.(2007){DiSanti}, {Anderson}, {Villanueva}, {Bonev},
  {Magee-Sauer}, {Gibb}, and {Mumma}}]{disa+:2007}
{DiSanti}, M.~A., {Anderson}, W.~M., {Villanueva}, G.~L., {Bonev}, B.~P.,
  {Magee-Sauer}, K., {Gibb}, E.~L., {Mumma}, M.~J., 2007. {Depleted carbon
  monoxide in fragment C of the Jupiter-family comet 73P/Schwassmann-Wachmann
  3}. \apjl 661, L101--L104.

\bibitem[{{DiSanti} and {Mumma}(2008)}]{disa-mumm:2008}
{DiSanti}, M.~A., {Mumma}, M.~J., 2008. {Reservoirs for comets: compositional
  differences based on infrared observations}. Space Scie. Rev. (in press).

\bibitem[{{Drahus} et~al.(2007{\natexlab{a}}){Drahus}, {Kueppers}, {Jarchow},
  {Paganini}, {Hartogh}, and {Villanueva}}]{drah+:2007}
{Drahus}, M., {Kueppers}, M., {Jarchow}, C., {Paganini}, L., {Hartogh}, P.,
  {Villanueva}, G.~L., 2007{\natexlab{a}}. {Submillimeter monitoring of the HCN
  molecule in fragment C of the split comet 73P/Schwassmann-Wachmann 3}. \baas
  39, 508.

\bibitem[{{Drahus} et~al.(2008{\natexlab{a}}){Drahus}, {Paganini}, {Ziurys},
  and {Peters}}]{drah+:2008-IAUC}
{Drahus}, M., {Paganini}, L., {Ziurys}, L., {Peters}, W., 2008{\natexlab{a}}.
  {Comet 17P/Holmes}. CBET 1289.

\bibitem[{{Drahus} et~al.(2008{\natexlab{b}}){Drahus}, {Paganini}, {Ziurys},
  {Peters}, {Jarchow}, and {Hartogh}}]{drah+:2008-ACM}
{Drahus}, M., {Paganini}, L., {Ziurys}, L., {Peters}, W., {Jarchow}, C.,
  {Hartogh}, P., 2008{\natexlab{b}}. {The recent mega-outburst of comet
  17P/Holmes at millimeter wavelengths}. In: Asteroids, Comets, Meteors 2008,
  Abstract \#8340, Lunar and Planetary Institute Contribution No 1405, Houston
  (CD-ROM).

\bibitem[{{Drahus} et~al.(2007{\natexlab{b}}){Drahus}, {Paganini}, {Ziurys},
  {Peters}, {Soukup}, and {Begam}}]{drah+:2007-IAUC}
{Drahus}, M., {Paganini}, L., {Ziurys}, L., {Peters}, W., {Soukup}, M.,
  {Begam}, M., 2007{\natexlab{b}}. {Comet 17P/Holmes}. \iaucirc 8891.

\bibitem[{{Fern\'andez}(2008)}]{fern:2008}
{Fern\'andez}, J.~A., 2008. {Origin of comet nuclei and dynamics}. Space Scie.
  Rev. (in press).

\bibitem[{{Festou} et~al.(2001){Festou}, {Gunnarsson}, {Rickman}, {Winnberg},
  and {Tancredi}}]{fest+:2001}
{Festou}, M.~C., {Gunnarsson}, M., {Rickman}, H., {Winnberg}, A., {Tancredi},
  G., 2001. {The activity of comet 29P/Schwassmann-Wachmann 1 monitored through
  its CO J=2--1 radio line}. Icarus 150, 140--150.

\bibitem[{{Fink}(2006)}]{fink:2006}
{Fink}, U., 2006. {How Tempel 1 fits into the ensemble of comets: a
  spectroscopic perspective}. In: {Sterken}, C., {K\"{a}ufl}, H.~U. (Eds.),
  Deep Impact as a world observatory event: synergies in space, time and
  wavelength (abstract book). Vrije Universiteit Brussel and ESO, p.~83.

\bibitem[{{Fray} et~al.(2005){Fray}, {B{\'e}nilan}, {Cottin}, {Gazeau}, and
  {Crovisier}}]{fray+:2005}
{Fray}, N., {B{\'e}nilan}, Y., {Cottin}, H., {Gazeau}, M.-C., {Crovisier}, J.,
  2005. {The origin of the CN radical in comets: A review from observations and
  models}. \planss 53, 1243--1262.

\bibitem[{{Fulle} et~al.(2004){Fulle}, {Barbieri}, {Cremonese}, {Rauer},
  {Weiler}, {Milani}, and {Ligustri}}]{full+:2004}
{Fulle}, M., {Barbieri}, C., {Cremonese}, G., {Rauer}, H., {Weiler}, M.,
  {Milani}, G., {Ligustri}, R., 2004. {The dust environment of comet
  67P/Churyumov-Gerasimenko}. \aap 422, 357--368.

\bibitem[{{Galt}(1987)}]{galt:1987}
{Galt}, J., 1987. {Observations of comet Giacobini-Zinner at the 1.667 GHz OH
  line}. \aj 94, 174--177.

\bibitem[{{G\'erard}(1990)}]{gera:1990}
{G\'erard}, E., 1990. {The discrepancy between OH production rates deduced from
  radio and ultraviolet observations of comets. I - A comparative study of OH
  radio and UV observations of P/Halley 1986 III in late November and early
  December 1985}. \aap 230, 489--503.

\bibitem[{{G\'erard} et~al.(1988){G\'erard}, {Bockel\'ee-Morvan}, {Bourgois},
  {Colom}, and {Crovisier}}]{gera+:1988}
{G\'erard}, E., {Bockel\'ee-Morvan}, D., {Bourgois}, G., {Colom}, P.,
  {Crovisier}, J., 1988. {Observations of the OH radio lines in comet
  P/Giacobini-Zinner 1985 XII}. \aaps 74, 485--495.

\bibitem[{{Gibson} and {Hobbs}(1981)}]{gibs-hobb:1981}
{Gibson}, D.~M., {Hobbs}, R.~W., 1981. {On the microwave emission from comets}.
  \apj 248, 863--866.

\bibitem[{{Green}(2007)}]{gree:2007}
{Green}, D.~W.~E., 2007. {Comet 17P/Holmes}. \iaucirc 8886.

\bibitem[{{Gulkis} et~al.(2007){Gulkis}, {Frerking}, {Crovisier}, {Beaudin},
  {Hartogh}, {Encrenaz}, {Koch}, {Kahn}, {Salinas}, {Nowicki}, {Irigoyen},
  {Janssen}, {Stek}, {Hofstadter}, {Allen}, {Backus}, {Kamp}, {Jarchow},
  {Steinmetz}, {Deschamps}, {Krieg}, {Gheudin}, {Bockel{\'e}e-Morvan}, {Biver},
  {Encrenaz}, {Despois}, {Ip}, {Lellouch}, {Mann}, {Muhleman}, {Rauer},
  {Schloerb}, and {Spilker}}]{gulk+:2007}
{Gulkis}, S., {Frerking}, M., {Crovisier}, J., {Beaudin}, G., {Hartogh}, P.,
  {Encrenaz}, P., {Koch}, T., {Kahn}, C., {Salinas}, Y., {Nowicki}, R.,
  {Irigoyen}, R., {Janssen}, M., {Stek}, P., {Hofstadter}, M., {Allen}, M.,
  {Backus}, C., {Kamp}, L., {Jarchow}, C., {Steinmetz}, E., {Deschamps}, A.,
  {Krieg}, J., {Gheudin}, M., {Bockel{\'e}e-Morvan}, D., {Biver}, N.,
  {Encrenaz}, T., {Despois}, D., {Ip}, W., {Lellouch}, E., {Mann}, I.,
  {Muhleman}, D., {Rauer}, H., {Schloerb}, P., {Spilker}, T., 2007. {MIRO:
  Microwave Instrument for Rosetta Orbiter}. Space Science Reviews 128,
  561--597.

\bibitem[{{Gunnarsson}(2003)}]{gunn:2003}
{Gunnarsson}, M., 2003. {Icy grains as a source of CO in comet
  29P/Schwassmann-Wachmann 1}. \aap 398, 353--361.

\bibitem[{{Gunnarsson} et~al.(2008){Gunnarsson}, {Bockel\'ee-Morvan}, {Biver},
  {Crovisier}, and {Rickman}}]{gunn+:2008}
{Gunnarsson}, M., {Bockel\'ee-Morvan}, D., {Biver}, N., {Crovisier}, J.,
  {Rickman}, H., 2008. {Mapping the carbon monoxide coma of comet
  29P/Schwassmann-Wachmann 1}. Astron. Astrophys. 484, 537--546.

\bibitem[{{Gunnarsson} et~al.(2002){Gunnarsson}, {Rickman}, {Festou},
  {Winnberg}, and {Tancredi}}]{gunn+:2002}
{Gunnarsson}, M., {Rickman}, H., {Festou}, M.~C., {Winnberg}, A., {Tancredi},
  G., 2002. {An extended CO source around comet 29P/Schwassmann-Wachmann 1}.
  Icarus 157, 309--322.

\bibitem[{{Harmon} et~al.(2005){Harmon}, {Nolan}, {Ostro}, and
  {Campbell}}]{harm+:2005}
{Harmon}, J.~K., {Nolan}, M.~C., {Ostro}, S.~J., {Campbell}, D.~B., 2005.
  {Radar studies of comet nuclei and grain comae}. In: {Festou}, M.~C.,
  {Keller}, H.~U., {Weaver}, H.~A. (Eds.), Comets II. Univ. Arizona Press,
  Tucson, pp. 265--279.

\bibitem[{{Howell} et~al.(2007{\natexlab{a}}){Howell}, {Lovell}, {Butler}, and
  {Schloerb}}]{howe+:2007a}
{Howell}, E.~S., {Lovell}, A.~J., {Butler}, B., {Schloerb}, F.~P.,
  2007{\natexlab{a}}. {Radio OH observations of 9P/Tempel 1 before and after
  Deep Impact}. Icarus 187, 228--239.

\bibitem[{{Howell} et~al.(2004){Howell}, {Lovell}, {Nolan}, and
  {Schloerb}}]{howe+:2004}
{Howell}, E.~S., {Lovell}, A.~J., {Nolan}, M.~C., {Schloerb}, F.~P., 2004.
  {Recent observations of the 18-cm OH lines in bright comets with the Arecibo
  ALFA receiver}. \baas 36, 1121.

\bibitem[{{Howell} et~al.(2007{\natexlab{b}}){Howell}, {Nolan}, {Harmon},
  {Lovell}, {Benner}, {Ostro}, {Campbell}, and {Margot}}]{howe+:2007b}
{Howell}, E.~S., {Nolan}, M.~C., {Harmon}, J.~K., {Lovell}, A.~J., {Benner},
  L.~A., {Ostro}, S.~J., {Campbell}, D.~B., {Margot}, J., 2007{\natexlab{b}}.
  {Radar and radio observations of the fragmented comet
  73P/Schwassmann-Wachmann 3}. \baas 39, 486.

\bibitem[{{Irvine} and {Schloerb}(2002)}]{irvi-schl:2002}
{Irvine}, W.~M., {Schloerb}, F.~P., 2002. {Observations of deuterated molecules
  with the Large Millimeter Telescope}. \planss 50, 1179--1184.

\bibitem[{{Irvine} and {Schloerb}(2005)}]{irvi-schl:2005}
{Irvine}, W.~M., {Schloerb}, F.~P., 2005. {The Large Millimeter Telescope- El
  Gran Telescopio Milimetrico}. \baas 37, 652.

\bibitem[{{Jehin} et~al.(2004){Jehin}, {Manfroid}, {Cochran}, {Arpigny},
  {Zucconi}, {Hutsem{\'e}kers}, {Cochran}, {Endl}, and {Schulz}}]{jehi+:2004}
{Jehin}, E., {Manfroid}, J., {Cochran}, A.~L., {Arpigny}, C., {Zucconi}, J.-M.,
  {Hutsem{\'e}kers}, D., {Cochran}, W.~D., {Endl}, M., {Schulz}, R., 2004. {The
  anomalous $^{14}$N/$^{15}$N ratio in comets 122P/1995 S1 (de Vico) and
  153P/2002 C1 (Ikeya-Zhang)}. \apjl 613, L161--L164.

\bibitem[{{Jewitt} et~al.(2008){Jewitt}, {Garland}, and {Aussel}}]{jewi+:2008}
{Jewitt}, D., {Garland}, C.~A., {Aussel}, H., 2008. {Deep search for carbon
  monoxide in cometary precursors using millimeter wave spectroscopy}. Astron.
  J. 135, 400--407.

\bibitem[{{Jewitt} and {Luu}(1990)}]{jewi-luu:1990}
{Jewitt}, D., {Luu}, J., 1990. {The submillimeter radio continuum of comet
  P/Brorsen-Metcalf}. \apj 365, 738--747.

\bibitem[{{Jewitt} and {Luu}(1992)}]{jewi-luu:1992}
{Jewitt}, D., {Luu}, J., 1992. {Submillimeter continuum emission from comets}.
  Icarus 100, 187--196.

\bibitem[{{Jewitt} and {Matthews}(1999)}]{jewi-matt:1999}
{Jewitt}, D., {Matthews}, H., 1999. {Particulate mass loss from comet
  Hale-Bopp}. \aj 117, 1056--1062.

\bibitem[{{Jewitt} et~al.(1997){Jewitt}, {Matthews}, {Owen}, and
  {Meier}}]{jewi+:1997}
{Jewitt}, D., {Matthews}, H.~E., {Owen}, T., {Meier}, R., 1997. {The 12C/13C,
  14N/15N and 32S/ 34S isotope ratios in comet Hale-Bopp (C/1995 O1).} Science
  278, 90--93.

\bibitem[{{Jewitt}(1996)}]{jewi:1996}
{Jewitt}, D.~C., 1996. {Debris From Comet P/Swift-Tuttle}. \aj 111, 1713--1717.

\bibitem[{{Jewitt} and {Matthews}(1997)}]{jewi-matt:1997}
{Jewitt}, D.~C., {Matthews}, H.~E., 1997. {Submillimeter continuum observations
  of comet Hyakutake (1196 B2)}. \aj 113, 1145--1151.

\bibitem[{{Jones} et~al.(2006){Jones}, {Sarkissian}, {Burton}, {Voronkov}, and
  {Filipovi{\'c}}}]{jone+:2006}
{Jones}, P.~A., {Sarkissian}, J.~M., {Burton}, M.~G., {Voronkov}, M.~A.,
  {Filipovi{\'c}}, M.~D., 2006. {Radio observations of comet 9P/Tempel 1 with
  the Australia Telescope facilities during the Deep Impact encounter}. \mnras
  369, 1995--2000.

\bibitem[{{Kobayashi} et~al.(2007){Kobayashi}, {Kawakita}, {Mumma}, {Bonev},
  {Watanabe}, and {Fuse}}]{koba+:2007}
{Kobayashi}, H., {Kawakita}, H., {Mumma}, M.~J., {Bonev}, B.~P., {Watanabe},
  J.-I., {Fuse}, T., 2007. {Organic volatiles in comet
  73P-B/Schwassmann-Wachmann 3 observed during its outburst: A clue to the
  formation region of the Jupiter-family comets}. \apjl 668, L75--L78.

\bibitem[{{Lecacheux} et~al.(2003){Lecacheux}, {Biver}, {Crovisier},
  {Bockel{\'e}e-Morvan}, {Baron}, {Booth}, {Encrenaz}, {Flor{\'e}n}, {Frisk},
  {Hjalmarson}, {Kwok}, {Mattila}, {Nordh}, {Olberg}, {Olofsson}, {Rickman},
  {Sandqvist}, {von Sch{\'e}ele}, {Serra}, {Torchinsky}, {Volk}, and
  {Winnberg}}]{leca+:2003}
{Lecacheux}, A., {Biver}, N., {Crovisier}, J., {Bockel{\'e}e-Morvan}, D.,
  {Baron}, P., {Booth}, R.~S., {Encrenaz}, P., {Flor{\'e}n}, H.-G., {Frisk},
  U., {Hjalmarson}, {\AA}., {Kwok}, S., {Mattila}, K., {Nordh}, L., {Olberg},
  M., {Olofsson}, A.~O.~H., {Rickman}, H., {Sandqvist}, A., {von Sch{\'e}ele},
  F., {Serra}, G., {Torchinsky}, S., {Volk}, K., {Winnberg}, A., 2003.
  {Observations of water in comets with Odin}. \aap 402, L55--L58.

\bibitem[{{Lellouch}(1996)}]{lell:1996}
{Lellouch}, E., 1996. {Chemistry induced by the impacts: observations}. In:
  {Noll}, K.~S., {Weaver}, H.~A., {Feldman}, P.~D. (Eds.), The Collision of
  Comet Shoemaker-Levy 9 and Jupiter. Cambridge University Press, pp. 213--242.

\bibitem[{{Levison}(1996)}]{levi:1996}
{Levison}, H.~F., 1996. {Comet taxonomy}. In: {Rettig}, T., {Hahn}, J.~M.
  (Eds.), ASP Conf. Ser. 107: Completing the Inventory of the Solar System. pp.
  173--191.

\bibitem[{{Lis} et~al.(2008){Lis}, {Bockel\'ee-Morvan}, {Boissier},
  {Crovisier}, {Biver}, and {Charnley}}]{lis+:2008}
{Lis}, D.~C., {Bockel\'ee-Morvan}, D., {Boissier}, J., {Crovisier}, J.,
  {Biver}, N., {Charnley}, S.~B., 2008. {Hydrogen isocyanide in comet
  73P/Schwassmann-Wachmann (fragment B)}. \apj 695, 931--936.

\bibitem[{{Lovell} et~al.(2006){Lovell}, {Howell}, {Marine}, {Butler}, and
  {Schloerb}}]{love+:2006}
{Lovell}, A.~J., {Howell}, E.~S., {Marine}, H., {Butler}, B.~J., {Schloerb},
  F.~P., 2006. {OH radio mapping observations of comet 73P/Schwassmann-Wachmann
  3}. \baas 38, 490.

\bibitem[{{Lovell} et~al.(2002){Lovell}, {Howell}, and {Schloerb}}]{love+:2002}
{Lovell}, A.~J., {Howell}, E.~S., {Schloerb}, F.~P., 2002. {Spectral line
  mapping observations of 18-cm OH lines in comets}. \baas 34, 869.

\bibitem[{{Lowry} et~al.(2008){Lowry}, {Fitzsimmons}, {Lamy}, and
  {Weissman}}]{lowr+:2008}
{Lowry}, S., {Fitzsimmons}, A., {Lamy}, P., {Weissman}, P., 2008. {Kuiper Belt
  Objects in the planetary region: the Juiter-family comets}. In: {Barucci},
  M.~A., {Boehnhardt}, H., {Cruikshank}, D.~P., {Morbidelli}, A. (Eds.), The
  Solar System beyond Neptune. Univ. Arizona Press, Tucson, pp. 397--410.

\bibitem[{{Manfroid} et~al.(2005){Manfroid}, {Jehin}, {Hutsem{\'e}kers},
  {Cochran}, {Zucconi}, {Arpigny}, {Schulz}, and {St{\"u}we}}]{manf+:2005}
{Manfroid}, J., {Jehin}, E., {Hutsem{\'e}kers}, D., {Cochran}, A., {Zucconi},
  J.-M., {Arpigny}, C., {Schulz}, R., {St{\"u}we}, J.~A., 2005. {Isotopic
  abundance of nitrogen and carbon in distant comets}. \aap 432, L5--L8.

\bibitem[{{Marsden}(2008)}]{mars:2008}
{Marsden}, B., 2008. {The obital properties of Jupiter Family Comets}. Planet.
  Space Scie. (this issue).

\bibitem[{{Meech} et~al.(2005){Meech}, {Ageorges}, {A'Hearn}, and {and a host
  of colleagues}}]{meec+:2005}
{Meech}, K.~J., {Ageorges}, N., {A'Hearn}, M.~F., {and a host of colleagues},
  2005. {Deep Impact: Observations from a worldwide Earth-based campaign}.
  Science 310, 265--269.

\bibitem[{{Milam} et~al.(2006){Milam}, {Apponi}, {Ziurys}, and
  {Wyckoff}}]{mila+:2006}
{Milam}, S.~N., {Apponi}, A.~J., {Ziurys}, L.~M., {Wyckoff}, S., 2006. {Comet
  73P/Schwassmann-Wachmann}. \iaucirc 8702.

\bibitem[{{Morbidelli}(2008)}]{morb:2008}
{Morbidelli}, A., 2008. {Comets and their reservoirs: Current dynamics and
  primodial evolution}. In: {Jewitt}, D., {Morbidelli}, A., {Rauer}, H. (Eds.),
  Trans-Neptunian Objects and Comets. Springer, Berlin, Heidelberg, pp.
  79--163.

\bibitem[{{Morbidelli} et~al.(2008){Morbidelli}, {Levison}, and
  {Gomes}}]{morb+:2008}
{Morbidelli}, A., {Levison}, H.~F., {Gomes}, R., 2008. {The dynamical structure
  of the Kuiper Belt and its primordial origin}. In: {Barucci}, M.~A.,
  {Boehnhardt}, H., {Cruikshank}, D.~P., {Morbidelli}, A. (Eds.), The Solar
  System beyond Neptune. Univ. Arizona Press, Tucson, pp. 275--292.

\bibitem[{{Moreno} et~al.(2003){Moreno}, {Marten}, {Matthews}, and
  {Biraud}}]{more+:2003}
{Moreno}, R., {Marten}, A., {Matthews}, H.~E., {Biraud}, Y., 2003. {Long-term
  evolution of CO, CS and HCN in Jupiter after the impacts of comet
  Shoemaker-Levy 9}. \planss 51, 591--611.

\bibitem[{{Mumma} et~al.(2002){Mumma}, {Disanti}, {Dello Russo}, {Magee-Sauer},
  {Gibb}, and {Novak}}]{mumm+:2002}
{Mumma}, M.~J., {Disanti}, M.~A., {Dello Russo}, N., {Magee-Sauer}, K., {Gibb},
  E., {Novak}, R., 2002. {The organic volatile composition of Oort cloud
  comets: evidence for chemical diversity in the giant-planets' nebular
  region}. In: {Warmbein}, B. (Ed.), ESA SP-500: Asteroids, Comets, and
  Meteors: ACM 2002. pp. 753--762.

\bibitem[{{Mumma} et~al.(2000){Mumma}, {DiSanti}, {Dello Russo}, {Magee-Sauer},
  and {Rettig}}]{mumm+:2000}
{Mumma}, M.~J., {DiSanti}, M.~A., {Dello Russo}, N., {Magee-Sauer}, K.,
  {Rettig}, T.~W., 2000. {Detection of CO and ethane in comet
  21P/Giacobini-Zinner: Evidence for variable chemistry in the outer Solar
  Nebula}. \apjl 531, L155--L159.

\bibitem[{{Neufeld} et~al.(2000){Neufeld}, {Stauffer}, {Bergin}, {Kleiner},
  {Patten}, {Wang}, {Ashby}, {Chin}, {Erickson}, {Goldsmith}, {Harwit}, {Howe},
  {Koch}, {Plume}, {Schieder}, {Snell}, {Tolls}, {Winnewisser}, {Zhang}, and
  {Melnick}}]{neuf+:2000}
{Neufeld}, D.~A., {Stauffer}, J.~R., {Bergin}, E.~A., {Kleiner}, S.~C.,
  {Patten}, B.~M., {Wang}, Z., {Ashby}, M.~L.~N., {Chin}, G., {Erickson},
  N.~R., {Goldsmith}, P.~F., {Harwit}, M., {Howe}, J.~E., {Koch}, D.~G.,
  {Plume}, R., {Schieder}, R., {Snell}, R.~L., {Tolls}, V., {Winnewisser}, G.,
  {Zhang}, Y.~F., {Melnick}, G.~J., 2000. {Submillimeter Wave Astronomy
  Satellite observations of water vapor toward comet C/1999 H1 (Lee)}. \apjl
  539, L151--L154.

\bibitem[{{Noll} et~al.(1996){Noll}, {Weaver}, and {Feldman}}]{noll+:1996}
{Noll}, K.~S., {Weaver}, H.~A., {Feldman}, P., 1996. {The Collision of Comet
  Shoemaker-Levy 9 and Jupiter}. Cambridge University Press.

\bibitem[{{Norris} et~al.(1985){Norris}, {Forster}, and {Duncan}}]{norr+:1985}
{Norris}, R.~P., {Forster}, J.~R., {Duncan}, R.~A., 1985. {Detection of OH
  maser emission from comet Giacobini-Zinner}. Proc. Astron. Soc. Australia 6,
  180--181.

\bibitem[{{Rauer} et~al.(1997){Rauer}, {Biver}, {Crovisier},
  {Bockel{\'e}e-Morvan}, {Colom}, {Despois}, {Ip}, {Jorda}, {Lellouch},
  {Paubert}, and {Thomas}}]{raue+:1997}
{Rauer}, H., {Biver}, N., {Crovisier}, J., {Bockel{\'e}e-Morvan}, D., {Colom},
  P., {Despois}, D., {Ip}, W.-H., {Jorda}, L., {Lellouch}, E., {Paubert}, G.,
  {Thomas}, N., 1997. {Millimetric and optical observations of Chiron}. \planss
  45, 799--805.

\bibitem[{{Schleicher} and {A'Hearn}(1988)}]{schl-ahea:1988}
{Schleicher}, D.~G., {A'Hearn}, M.~F., 1988. {The fluorescence of cometary OH}.
  \apj 331, 1058--1077.

\bibitem[{{Schloerb}(1988)}]{schl:1988}
{Schloerb}, F.~P., 1988. {Collisional quenching of cometary emission in the 18
  centimeter OH transitions}. \apj 332, 524--530.

\bibitem[{{Sekanina}(2005)}]{seka:2005}
{Sekanina}, Z., 2005. {Comet 73P/Schwassmann-Wachmann: nucleus fragmentation,
  its light-curve signature, and close approach to Earth in 2006}. Int. Comet
  Quarterly 27, 225--240.

\bibitem[{{Senay} and {Jewitt}(1994)}]{sena-jewi:1994}
{Senay}, M.~C., {Jewitt}, D., 1994. {Coma formation driven by carbon-monoxide
  release from comet Schwassmann-Wachmann 1}. \nat 371, 229--231.

\bibitem[{{Snyder}(1982)}]{snyd:1982}
{Snyder}, L.~E., 1982. {A review of radio observations of comets}. Icarus 51,
  1--24.

\bibitem[{{Soderblom} et~al.(2002){Soderblom}, {Becker}, {Bennett}, {Boice},
  {Britt}, {Brown}, {Buratti}, {Isbell}, {Giese}, {Hare}, {Hicks},
  {Howington-Kraus}, {Kirk}, {Lee}, {Nelson}, {Oberst}, {Owen}, {Rayman},
  {Sandel}, {Stern}, {Thomas}, and {Yelle}}]{sode+:2002}
{Soderblom}, L.~A., {Becker}, T.~L., {Bennett}, G., {Boice}, D.~C., {Britt},
  D.~T., {Brown}, R.~H., {Buratti}, B.~J., {Isbell}, C., {Giese}, B., {Hare},
  T., {Hicks}, M.~D., {Howington-Kraus}, E., {Kirk}, R.~L., {Lee}, M.,
  {Nelson}, R.~M., {Oberst}, J., {Owen}, T.~C., {Rayman}, M.~D., {Sandel},
  B.~R., {Stern}, S.~A., {Thomas}, N., {Yelle}, R.~V., 2002. {Observations of
  comet 19P/Borrelly by the Miniature Integrated Camera and Spectrometer aboard
  Deep Space 1}. Science 296, 1087--1091.

\bibitem[{{Tacconi-Garman} et~al.(1990){Tacconi-Garman}, {Schloerb}, and
  {Claussen}}]{tacc+:1990}
{Tacconi-Garman}, L.~E., {Schloerb}, F.~P., {Claussen}, M.~J., 1990. {High
  spectral resolution observations and kinematic modeling of the 1667 MHz
  hyperfine transition of OH in Comets Halley (1982i), Giacobini-Zinner
  (1984e), Hartley-Good (1985l), Thiele (1985m), and Wilson (1986l)}. \apj 364,
  672--686.

\bibitem[{{Tozzi} et~al.(2007){Tozzi}, {Palagi}, {Codella}, {Poppi}, and
  {Crovisier}}]{tozz+:2007}
{Tozzi}, G.~P., {Palagi}, F., {Codella}, C., {Poppi}, S., {Crovisier}, J.,
  2007. {Search for ammonia radio emission in comet 9P/Tempel 1 after the Deep
  Impact event}. In: {Sterken}, C., {K\"{a}ufl}, H.~U. (Eds.), Deep Impact as a
  world observatory event: synergies in space, time and wavelength.
  Springer-Verlag, ESO Astrophysics Symposia, in press.

\bibitem[{{Turner}(1974)}]{turn:1974}
{Turner}, B.~E., 1974. {Detection of OH at 18-centimeter wavelength in comet
  Kohoutek (1973f)}. \apjl 189, L137--L139.

\bibitem[{{Villanueva} et~al.(2006){Villanueva}, {Mumma}, {Bonev}, {Drahus},
  {Paganini}, {Kueppers}, {DiSanti}, {Magee-Sauer}, {Hartogh}, {Jarchow},
  {Milam}, and {Ziurys}}]{vill+:2006-BAAS}
{Villanueva}, G., {Mumma}, M.~J., {Bonev}, B.~P., {Drahus}, M., {Paganini}, L.,
  {Kueppers}, M., {DiSanti}, M.~A., {Magee-Sauer}, K., {Hartogh}, P.,
  {Jarchow}, C., {Milam}, S., {Ziurys}, L.~M., 2006. {Submillimeter molecular
  observations of the split ecliptic comet 73P/Schwassman-Wachmann 3}. \baas
  38, 485.

\bibitem[{{Weaver} et~al.(1999){Weaver}, {Chin}, {Bockel{\'e}e-Morvan},
  {Crovisier}, {Brooke}, {Cruikshank}, {Geballe}, {Kim}, and
  {Meier}}]{weav+:1999}
{Weaver}, H.~A., {Chin}, G., {Bockel{\'e}e-Morvan}, D., {Crovisier}, J.,
  {Brooke}, T.~Y., {Cruikshank}, D.~P., {Geballe}, T.~R., {Kim}, S.~J.,
  {Meier}, R., 1999. {An infrared investigation of volatiles in comet
  21P/Giacobini-Zinner}. Icarus 142, 482--497.

\bibitem[{{Webber} and {Snyder}(1977)}]{webb-snyd:1977}
{Webber}, J.~C., {Snyder}, L.~E., 1977. {Detection of radio OH in periodic
  comet d'Arrest}. \apjl 214, L45--L46.

\bibitem[{{Webber} et~al.(1977){Webber}, {Snyder}, and {Ensinger}}]{webb+:1977}
{Webber}, J.~C., {Snyder}, L.~E., {Ensinger}, J., 1977. {Observations of radio
  OH in periodic comet Encke.} \baas 9, 564.

\bibitem[{{Womack} and {Stern}(1995)}]{woma-ster:1995}
{Womack}, M., {Stern}, S.~A., 1995. {(2060) Chiron = comet 95P/Chiron}.
  \iaucirc 6193.

\bibitem[{{Womack} and {Stern}(1999)}]{woma-ster:1999}
{Womack}, M., {Stern}, S.~A., 1999. {The detection of carbon monoxide gas
  emission in (2060) Chiron (in Russian)}. Astron. Vest. 33, 216--221.

\bibitem[{{Woodney} et~al.(2004){Woodney}, {Fernandez}, and
  {Owen}}]{wood+:2004}
{Woodney}, L.~M., {Fernandez}, Y.~R., {Owen}, T.~C., 2004. {Measuring HCN in
  comet 2P/Encke}. \baas 36, 1146.

\bibitem[{{Woodney} et~al.(2003){Woodney}, {Owen}, and
  {Fernandez}}]{wood+:2003}
{Woodney}, L.~M., {Owen}, T.~C., {Fernandez}, Y.~R., 2003. {Comet 2P/Encke}.
  \iaucirc 8239.

\bibitem[{{Zakharov} et~al.(2007){Zakharov}, {Bockel{\'e}e-Morvan}, {Biver},
  {Crovisier}, and {Lecacheux}}]{zakh+:2007}
{Zakharov}, V., {Bockel{\'e}e-Morvan}, D., {Biver}, N., {Crovisier}, J.,
  {Lecacheux}, A., 2007. {Radiative transfer simulation of water rotational
  excitation in comets. Comparison of the Monte Carlo and escape probability
  methods}. \aap 473, 303--310.

\bibitem[{{Ziurys} et~al.(1999){Ziurys}, {Savage}, {Brewster}, {Apponi},
  {Pesch}, and {Wyckoff}}]{ziur+:1999}
{Ziurys}, L.~M., {Savage}, C., {Brewster}, M.~A., {Apponi}, A.~J., {Pesch},
  T.~C., {Wyckoff}, S., 1999. {Cyanide chemistry in comet Hale-Bopp (C/1995
  O1)}. \apjl 527, L67--L71.

\end{thebibliography}

}

\end{document}